\documentclass[manuscript]{aastex}

\usepackage{natbib}

\begin{document}

\footnotetext[1]{Last updated: \today}

\title{ROTATIONALLY TIME-RESOLVED VIS-SPECTROSCOPY OF (3200) PHAETHON}

\author{Daisuke Kinoshita}
\affil{Institute of Astronomy, National Central University, 300 Jhongda Rd., Chungli, Taoyuan, 320, Taiwan}

\author{Katsuhito Ohtsuka}
\affil{Tokyo Meteor Network, Daisawa 1--27--5, Setagaya, Tokyo, 155--0032, Japan}
\email{ohtsuka@jb3.so-net.ne.jp}

\author{Takashi Ito}
\affil{Center for Computational Astrophysics, National Astronomical Observatory of Japan, Osawa 2--21--1, Mitaka, Tokyo, 181--8588, Japan}

\author{Seidai Miyasaka}
\affil{Tokyo Metropolitan Government, Nishi-Shinjuku 2--8--1, Shinjuku, Tokyo, 163--8001, Japan}

\author{Tomoki Nakamura}
\affil{Department of Earth science, Graduate School of Science, Tohoku University, 6--3, Aramaki Aza Aoba, Aoba, Sendai, Miyagi 980--8578, Japan}

\author{Shinsuke Abe}
\affil{Department of Aerospace Engineering, Nihon University, 7--24--1, Narashinodai, Funabashi, Chiba 274--8501, Japan}

\author{Wen-Ping Chen}
\affil{Institute of Astronomy, National Central University, 300 Jhongda Rd., Chungli, Taoyuan, 320, Taiwan}

\begin{abstract}
Apollo-type near-Earth asteroid (3200) Phaethon, 
classified at the B/F-type taxonomy, 
probably the main mass of the Phaethon-Geminid stream complex (PGC), 
can be the most metamorphic C-complex asteroid in our solar system, 
since it is heated up to $\sim1000$ K by the solar radiation heating 
around its perihelion passages. 
Hence, the surface material of the asteroid may be easily decomposed 
in near-sun environment. 
Phaethon's spectrum exhibits extremely blue-slope in the VIS-NIR region
(so-called Phaethon Blue). 
Another candidate large member of the PGC, 
Apollo asteroid (155140) 2005 UD, shows a B/F-type color, 
however with a C-type-like red color over its $\sim 1/4$ rotational part, 
which implies an exposition of less metamorphosed primordial internal structure of 
the PGC precursor by a splitting or breakup event long ago. 
If so, some rotational part of Phaethon should show the C-type color 
as well as 2005 UD. 
Hence, we carried out the time-series VIS-spectroscopic observations of 
Phaethon in 30 November-- 2 December, 2007, 
using 1-m telescope at Lulin observatory in Taiwan 
in order to detect such a signature on Phaethon. 
Also, $R$-band photometries were simultaneously performed in order to complement
our spectroscopy. 
Consequently, we obtained a total of 68 VIS-spectrophotometric data, 
78\% of which show the B-type blue-color, 
as against the rest of 22\% showing the C-type red-color, 
as we expected.  
We successfully acquired rotationally time-resolved spectroscopic data, 
of which particular rotational phase shows a red-spectral slope as the C-type color, 
as 2005 UD does, suggesting longitudinal inhomogeneity on Phaethon's surface.  
We constrained this C-type red-colored area in the 
mid-latitude in Phaethon's southern hemisphere 
based on the rotationally time-resolved spectroscopy 
along with Phaethon's axial rotation state,   
of which size suggests the impact-induced origin of the PGC. 
We also surveyed the meteoritic analog of Phaethon's surface blue-color, 
and found thermally metamorphosed CI/CM chondrites 
as likely candidates.   
\end{abstract}

\keywords{minor planets, asteroids: individual (3200 Phaethon, 155140 2005 UD) ---
meteorites, meteors, meteoroids --- techniques: spectroscopic, photometric --- methods: observational --- Planetary systems}

\section{Introduction\label{sec:introduction}}
Near-Earth asteroids (NEAs) come around the inner solar system
from various source regions, bringing materials and information
that serve as important clues to understand the origin and
evolution of the solar system environment. Some of them 
may be extinct or dormant cometary nuclei, 
coming from the Jupiter family comet (JFC) region.   
Apollo-type NEA (3200) Phaethon \citep[= 1983 TB:][]{green1983}
has been well-known as one of the most enigmatic small bodies in the solar system, 
since it involves the Phaethon-Geminid stream complex 
\citep[hereafter, PGC:][]{whipple1983,babadzhanov1992,kasuga2009} 
even though Phaethon is not a comet but an asteroid.
So, the asteroid has long been regarded as an extinct cometary nucleus. 
However, the recent dynamical study by \citet{deleon2010} revealed that 
Phaethon was possibly chipped out from Main-belt asteroid (2) Pallas within 
$\sim30$ Myr, i.e., within a typical dynamical lifetime of NEAs 
\citep{gladman1997,ito2010}. 
This fact practically denies a cometary origin of Phaethon. 
The asteroid is not only an NEA but also a near-Sun asteroid (NSA).    
Its eccentricity $e$, semimajor axis $a$, and perihelion distance $q$ are 
respectively $e=0.89$, $a=1.27$ AU, and $q=0.14$ AU at the present evolutionary epoch, 
thus the relativistic effects of the Sun is not negligible 
when we study long-term dynamical evolution of Phaethon \citep{galushina2015}: 
more effective than the Yarkovsky effect on the asteroid.   
Since Phaethon's orbital plane is moderately inclined to the ecliptic plane at inclination, $i\sim 22^{\circ}$, 
the $q$-$i$ oscillating (and $e$ at antiphase) secular evolution is controlled under 
the Lidov--Kozai circulation mechanism with the $\omega$-cycle of $\sim 37,000$ yr \citep{urakawa2014}. 
 
Dynamical and physical behaviors of NSAs and small solar system bodies (hereafter, SSSB(s)) in 
near-sun environments have recently drown our particular interest \citep{jewitt2013a}. 
Phaethon is currently the largest object, $D \simeq 5.1$ km in size \citep{harris1998,hanus2016a}, 
among all the known NSAs.
The asteroid should suffer strong solar radiation heating around its perihelion passages,  
where sub-solar surface temperature of Phaethon is estimated to be at  
$800$--$1100$ K \citep{ohtsuka2009} 
from FRM \citep[Fast Rotating Model:][]{lebofsky1989} and NEATM \citep[Near-Earth Asteroid Thermal Model:][]{harris1998}.
This temperature is high enough to metamorphose primitive chondritic materials that 
have never undergone high-temperature heating in the solar-system history. 
  
Phaethon shows spectroscopically bluer than the solar color, 
as against usual C-type spectroscopy generally being in neutral--red
with the meteoritic analog of CI/CM chondrites. 
The visible (hereafter, VIS) to near infrared (hereafter, NIR) 
spectroscopic continuum of Phaethon indicates
a negative slope of normalized gradient of reflectivity  
along with the UV/$B$-band falling-off and lacking the 0.7-$\micron$ absorption feature due to 
decomposition of phyllosilicates. 
Thus the asteroid is classified at the B/F-type taxonomy as well as Pallas 
by the VIS--NIR spectrophotometries 
\citep[e.g.,][and others]{tholen1985,luu1990,binzel2001,licandro2007}, 
i.e., suggesting a dehydrated and metamorphosed C-type asteroid. 
In fact, Phaethon is a low-albedo asteroid, geometric albedo $p$v$=0.122 \pm 0.008$ 
along with $D = 5.1 \pm 0.2$ km-size and 
relatively high thermal inertia $\Gamma=600 \pm 200$ J m$^{-2}$s$^{-1/2}$ K$^{-1}$ 
\citep{hanus2016a,hanus2016b} 
almost equivalent to Cold Bokkeveld (CM2) of $<650$ J m$^{-2}$s$^{-1/2}$ K$^{-1}$ 
\citep{opeil2010}.  
These physical parameters are of typical B-type asteroids \citep{Ali-Lagoa2013}.
However, a strikingly matching meteoritic (surface) analog for Phaethon has never been identified yet.  
Nevertheless, thermally metamorphosed CI/CM chondrites 
are the best candidate analog for the C/G/B/F-type asteroids \citep{hiroi1993,hiroi1996}, 
since those carbonaceous chondrites underwent thermal history of high-temperature heating at
more than several $100$ K and dehydration for a certain period of time after 
aqueous alteration in their parent bodies \citep{nakamura2005,nakamura2006,nakato2008}. 
Indeed, Phaethon's sub-solar surface temperature of $800$--$1100$ K at perihelion 
is obviously higher than the temperatures  
starting serpentine-phyllosilicate decomposition  
at $\sim 573$ K and dehydration at $\sim 673$ K \citep{akai1992}. 
Serpentine in carbonaceous chondrites 
also show similar decomposition temperature of serpentine \citep{nozaki2006}.
Also, the temperatures starting saponite-phyllosilicate decomposition and dehydration 
are higher by $\sim 200$ K than those of serpentine-phyllosilicate \citep{nozaki2006}. 
Major constituents of CI and CM chondrites are saponite-rich and serpentine-rich mineralogy 
in their fine-grained matrix-phyllosilicates, respectively.  
In addition, the reflectance spectra of CI/CM chondrites show bluer-slope in the case of 
their grains being coarser and larger than 100 $\micron$-order in size \citep{cloutis2013}, 
so-called, ``grain size effects''.   
However, \citet{clark2010} advocated CK chondrites being another candidate analog 
for Phaethon, comparing their VIS--NIR spectral shapes.
Also, some Ureilites exhibit a blue spectral slope \citep{jenniskens2009,cloutis2010}, 
along with shallow Band I/$\rm I\!I$ (of mafic iron silicate absorption band) features,   
although Band I/$\rm I\!I$ features are generally not recognized in Phaethon's spectrum.  
Besides, space-weathering effects should strongly influence on the C-type asteroid spectroscopy, 
which tends to be darker and bluer again \citep{moroz2004,matsuoka2015}. 
\citet{dundon2005} found a spectral similarity between Phaethon and carbon black 
in the VIS--NIR range.
This makes us imagine that Phaethon's surface is coated by some ``carbonized" material produced by  
space-weathering \citep{moroz2004}. 
Phaethon would have originally been holding Pallas's surface and/or inner color,
if Pallas is actually the parent body of Phaethon. 
However, the negative spectral slope of Phaethon is extremely steeper in the VIS-range
by a factor of 40 than that of Pallas \citep{clark2010}. 
In fact, Phaethon is known to be considerably bluer object 
than any other SSSB, 
according to \citet{luu1990} and \citet{clark2010}. 
The physical parameters of Phaethon are bound up by G.~Hahn 
\footnote{http://earn.dlr.de/nea/003200.htm}. 
 
Another hot topic worth noting is that Phaethon is indeed an active asteroid 
as categorized by rocky comet \citep{jewitt2015}. 
Although no trace of Phaethon's cometary activities has been found yet 
by ground-based observations 
\citep[e.g.,][and others]{chamberlin1996,urakawa2002,hsieh2005}, 
recurrent weak cometary activities were 
detected in 2009, 2010, 2012, and 2016 by 
the spaceborne coronagraphic observatory, 
the Heliospheric Imager, HI-1/SECCHI onboard NASA's STEREO-A spacecraft   
\citep{jewitt2010,li2013,jewitt2013b,hui2017}. 
Then Phaethon's brightness increased by $\sim 2$ magnitude (hereafter, mag) near its perihelion.  
The dust tail was also found in 2009, 2012, and 2016   
whose $\beta$-syndyne indicated that released dust particles were 
in an effective radius of $\sim 1$ $\micron$-order:   
thus too small in size to replenish the Geminid meteoroid stream \citep{jewitt2013b,hui2017}. 
Those activities may be induced by dust ejections due to 
thermal fracture and/or thermal decomposition of surface materials around perihelion 
\citep{li2013,jewitt2013b}. 
Especially, the north-pole (NP) region of Phaethon will be selectively and successively 
heated over $1000$ K just after the perihelion passage, 
particularly due to non-rotational NEATM-like effect \citep{galushina2015}, 
if we take Phaethon's pole axis solution of \citet{krugly2002} or \citet{ansdell2014}:  
both pole solutions were determined to be high-obliquity close to Phaethon's orbital plane and the ecliptic plane. 
On the other hand, \citet{hanus2016a} very recently presented somewhat different pole axis solution 
with moderate-obliquity.   
This pole solution indicates that Phaethon's entire surface has been almost equally burnt out by the solar radiation heating 
rather than selective and successive NP-region heating,  
because its pole orientation should be substantially moved by secular pole axis precession.
If it is true, this may result in a saturation of the thermal metamorphism in Phaethon's entire surface. 
Nevertheless, if the PGC members (hereafter, PGCs) were generated by
any impact-induced splitting or breakup event on Phaethon (or PGC precursor), C-type-like surface,
implying the primordial internal structure of Phaethon (or PGC precursor), 
may be still exposed somewhere on Phaethon's surface.
If so, this C-type-like surface materials possibly consist of water- and volatile-bearing CI/CM-like 
primitive carbonaceous chondrites, which may contribute the cometary activity of Phaethon.     

In fact, Phaethon is probably the main mass among the PGCs, whereas  
the second-largest mass candidate is an 1.3-km-sized object, 
Apollo-type NEA (155140) 2005 UD \citep{ohtsuka2006}, which is also an NSA. 
Another large PGC candidate is Apollo-type NEA (225416) 1999 YC \citep{ohtsuka2008,kasuga2008}, 
though its orbital energy, $a^{-1}$, is somewhat smaller than Phaethon and 2005 UD. 
The dynamical motion of 2005 UD shows comparable $q$-$i$ oscillating Lidov--Kozai cycle 
with that of Phaethon, time-lagging by $\sim 4600$ yr \citep{ohtsuka2006}. 
From broadband photometric measurements, the surface color of 2005 UD indicates 
that it is most likely to have the B/F-type taxonomy, compared with that of Phaethon 
\citep{jewitt2006,kinoshita2007,kasuga2008}.   
The B/F-type asteroids are very rare, comprising only $\sim5\%$, among all the classified NEAs 
\citep{jewitt2006}: thus, the probability that Phaethon and 2005 UD would be found by randomly
drawing from all of the NEAs is extremely low.  
It is interesting that $\sim 1/4$ rotational part of 2005 UD 
shows a red-spectral slope, suggesting likely the C-type color \citep{kinoshita2007}.  
Hence, this $\sim 1/4$ part may imply an exposition of the primordial internal structure of 
the PGC precursor by a splitting or breakup event, 
i.e., less-metamorphic and  less space-weathered surface on 2005 UD. 
So, we hypothetically suspect that 
2005 UD have arisen from the disintegration of Phaethon (or PGC precursor) $\sim 1$ Myr ago \citep{ohtsuka2006}, 
a long time after Phaethon (or PGC precursor) being released from Pallas within 30 Myr ago, 
if the precursor actually attribute to Pallas.  
Then the PGCs would be generated, 
in which smaller decameter- and meter-sized bodies,  
far-smaller fragments and dusts may have been produced by power-law scaling.     
So, some decameter- and meter-sized PGC bodies may penetrate into the earth's atmosphere 
with relatively high velocity of $\sim 33$ km sec$^{-1}$, 
when most of them should be highly ablated and thus not survive.  
However a little bit of these bodies may survive and 
be preserved on the Earth's ground somewhere  
as the Phaethon (or PGC) meteorites \citep{madiedo2013}.  
The rest of fragments and dusts may generate 
the Geminid and Daytime Sextantid meteor streams and other PGC streams \citep{babadzhanov1992,ohtsuka1997}. 
Also it is slightly possible that NEA 2011 XA$_{3}$ is an impactor or an interloper of the PGC \citep{urakawa2014}. 

If our working hypothetical scenario is true, 
some rotational part of Phaethon should show the neutral- to red-spectral slope as the C-type color 
as 2005 UD does. 
This should be a strong signature of the splitting or breakup event on Phaethon (or PGC precursor) in the past,
and provide yet another strong piece of
evidence for a common origin of Phaethon and 2005 UD. 
However, the (past) other works for the VIS(--NIR) spectroscopic observations of Phaethon 
have been carried out by single frame shot or so. 
Hence, the existence of the C-type color on Phaethon's surface 
has been not well recognized, 
although a possible surface inhomogeneity was briefly discussed \citep{licandro2007, ohtsuka2009}. 

Therefore, to ensure our hypothesis, we carried out time-series
VIS spectroscopic observations of Phaethon in 30 November--2 December, 2007, 
covering more than one full rotational phase \citep[or period $P \sim 3.6$ hr: e.g.,][]{hanus2016a}, 
when the asteroid approached the Earth within $\sim 0.2$ AU 
and then brightened up to visual magnitude, $m_{v} \sim 13$ mag.
Also, $R$-band photometric observations  
were simultaneously performed in order to complement our spectroscopy. 
Detail of our observation is described in Section~\ref{sec:observation}.
We present the results of our rotationally time-resolved VIS-spectrophotometries of Phaethon 
and give discussions about them in Section~\ref{sec:result}.
Finally we summarize and conclude our works in Section~\ref{sec:conclusion}.

\section{Observations and Data reduction\label{sec:observation}}

\subsection{Spectroscopy and its testing\label{ssec:spectroscopy}}

The VIS spectroscopic observations of Phaethon
were carried out by using the Lulin 1.0-m (One-meteor) Telescope 
(thus commonly called ``LOT'')  
with a low-dispersion spectrograph named ``Hiyoyu'' at Lulin observatory 
operated by the Institute of Astronomy, National Central University.
The observatory is located near the summit of Mt. Lulin 
(MPC Observatory code: D35, $120^\circ 52' 25''$E, $23^\circ 28'7''$N, H$=2862$ m above sea level) 
in the central region of Taiwan. 
The spectrograph, Hiyoyu, which means rainbow in the language of native Taiwanese of Tsou tribe,
was mounted on the Cassegrain focus (F/8) of LOT and used for this study. 
Hiyoyu is almost identical to the Gunma Compact Spectrograph
(GCS) at the Gunma Astronomical Observatory \citep{hamane2002}. 
GCS was originally designed for F/12 of the Gunma telescope, 
so we attached a conversion lens to convert from F/8 into F/12 
before feeding the light into Hiyoyu for using LOT. 
The focal lengths of the collimating optics and camera optics are 
240 mm and 200 mm, respectively. 
The dispersion element is due to reflective gratings, 
which is replaceable for two modes for our instrument 
by selecting low resolution gratings of 300 gr mm$^{-1}$ or 
1200 gr mm$^{-1}$. 
Here we chose the grating of 300 gr mm$^{-1}$ along with the $390 \times 1.5$ arcsec slit 
to record the spectral shape along with overall features of Phaethon in the VIS-range. 
This 1.5-arcsec slit is in fact narrow in width for the VIS-spectroscopic observations.
However, since the seeing size of stellar FWHM
(full width at half maximum) at Lulin observatory is generally within 1.5 arcsec, 
the 1.5-arcsec width slit is acceptable for majority of uses. 
We used the Apogee 1K-CCD camera, AP-8 (Kodak KAF-1001E), as a detector for spectral images, 
whose format having $1024 \times 1024$ pixels (24 $\micron$ pixel$^{-1}$).  
Thus CCD chip size is $24.6 \times 24.6$ mm, which results in pixel scale being 0.62 arcsec pixel$^{-1}$.   
Resultant spectral resolution $R(\lambda/\Delta \lambda)$ is $\sim 330$ (at $\lambda = 5000$ \AA). 
The cooling temperature of the CCD is set to $-50^\circ$C, 
and this is achieved by both thermoelectric cooling and water circulation.
The spectral wavelength ranges between $3800$--$7600$ \AA, covering over the VIS-range. 
The SBIG Astronomical Instruments CCD camera, ST-8, was used as a slit viewer
monitoring camera, and was also used to adjust the telescope tracking manually
during the exposures of Phaethon. 
Although LOT has a function to perform the differential tracking, 
our main target, Phaethon, was in fact a fast moving NEA that quickly approaches the Earth  
and thus fine-adjustment of manual tracking was important 
to obtain high quality spectra. 
The Fe-Ne-Ar hollow cathode lamp is used for the wavelength calibration.
The specification of Hiyoyu
\footnote{for more detail, see http://www.lulin.ncu.edu.tw/lot/Hiyuyo\_spectrograph\_manual.pdf} 
is listed in Table~\ref{hiyoyu}.
 
In this way, the observations were made for three successive nights in 
30 November--2 December in 2007. 
We continuously took a 300-sec integration in each exposure  
when Phaethon has good visibility in observation. 
Consequently, a total of 68 VIS-spectroscopic data of Phaethon were successfully recorded, 
which may cover more than one full rotational phase, 
among which 24, 27, and 17 data were obtained in each night, respectively.
The log of the observations is summarized in Table~\ref{obslog}. 
Then we selected HR1544 and HR3454 as the spectrophotometric standard stars, 
and HIP22349 and HIP46404 as the solar-analog G2V stars. 
Their images were generally recorded every 30-minutes interval between observations of Phaethon. 
Sometimes we observed the Moon as another solar-analog star for reference. 
The Fe-Ne-Ar lamp spectra were also obtained along with the exposures. 
The dark and dome flat-field frames were taken before and after the observations of each night. 

All the spectral images were dark-subtracted and flat-fielded 
along the standard data reduction procedure 
using the image-analysis software package 
NOAO IRAF (Image Reduction and Analysis Facility). 
Extraction of the spectrum from two-dimensional spectral images, 
subtraction of the sky background, and wavelength and flux calibrations
were carried out by using the {\tt twodspec} and {\tt onedspec} tasks in IRAF.  
After that, an asteroid spectrum was divided by the spectra of the G2V star, 
and finally we obtained a scaled reflectance of the asteroid. 

\begin{figure}[htbp]
 \epsscale{0.7}
 \plotone{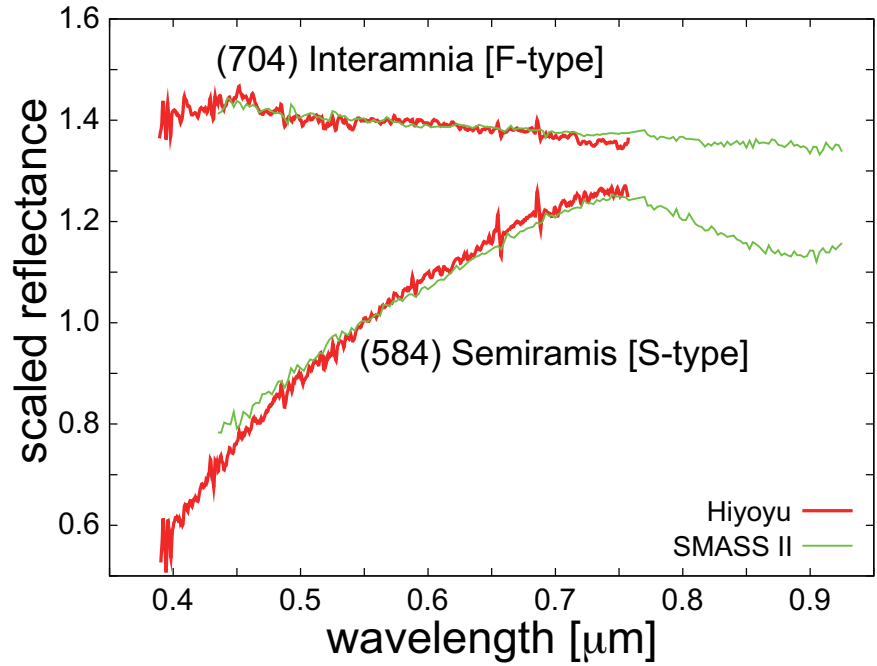}
 \caption{The VIS-spectra of (548) Semiramis and (704) Interamnia
 obtained by Hiyoyu, together with those from  
 the SMASS $\rm I\!I$ database for comparison. 
 They are normalized at 0.55 $\micron$.  
 The spectra of Interamnia are shifted by $+0.4$ in vertical direction. 
 See Section~\ref{ssec:spectroscopy} for more detail.}
 \label{known_asteroids}
\end{figure}

Before observing Phaethon each night, 
we performed the VIS-spectroscopic observations for 
some bright main belt asteroids to test and confirm the instrument
functionality, procedure of our observations,   
and subsequent calibration and data reduction process for Phaethon.
Then we obtained the VIS-spectra of 
Main-belt S-type asteroid (548) Semiramis and F-type asteroid (704) Interamnia,
as represent in Figure~\ref{known_asteroids}. 
Both the asteroid spectra using Hiyoyu show typical taxonomic features, respectively.   
The VIS-spectra of both the asteroids were in fact  
recorded by the SMASS $\rm I\!I$ 
\citep[Phase $\rm I\!I$ of the Small Main-Belt Asteroid Spectroscopic Survey:][]{bus2002a}, 
which are superimposed in Figure \ref{known_asteroids} for comparison. 
The SMASS  $\rm I\!I$ works are well-known 
as building global reputation in the VIS-NIR spectroscopy of asteroids.  
The first goal of our study is to obtain and check out the overall flux curve and shapes of our VIS-spectra. 
As can be seen from Figure~\ref{known_asteroids},  
the spectral flux curves for Semiramis and Interamnia, 
recorded by our Hiyoyu system, agree rather well with those by the SMASS $\rm I\!I$ works.
Hence, we can say that we have confirmed that the Hiyoyu system is safe
to study the VIS-spectroscopy of the SSSBs.


\subsection{Photometry\label{ssec:photometry}}

Simultaneous photometric observations for the target asteroid are important to 
complement our rotationally time-resolved spectroscopy. 
So, we can identify a rotational phase at the time for each spectrophotometry 
based on the rotational phase curve, i.e., lightcurve, constructed from time-series photometries. 
Such complementary photometric observations are carried out 
by using the 0.36-m Ritchey-Chretien (F/8) telescope at the Miyasaka observatory  
(MPC Observatory code: 366, $138^\circ 17'50''$E, $35^\circ 51'57''$N, H$=860$ m above sea level) in Japan. 
Then the SBIG 1K-CCD camera, STL-1001E (Kodak KAF-1001E), is used as a detector. 
This camera has $1024 \times 1024$ pixels (24 $\micron$ pixel$^{-1}$),   
thus the CCD chip size is $24.6 \times 24.6$ mm. 
So, the field of view (FOV) for the CCD chip is the square of $29'.4$ 
with a pixel scale of 1.72 arcsec pixel$^{-1}$. 
This is spatially well-sampled at the typical seeing size of stellar FWHM 
at the observatory, since, in winter season, the seeing conditions at 
observatories in east-Japan located near the coast of the Pacific ocean 
are generally not good.  
The stellar FWHMs are in fact 6--7 arcsec during our photometries.  
The CCD is cooled by the thermoelectric cooling, 
and the operation temperature of the CCD was $-35^\circ$C. 

In such a manner, $R$-band photometric observations of Phaethon 
were performed on two nights, 30 November and 1 December, 2007, 
simultaneously with the VIS-spectroscopic observations at Lulin. 
We took continuously a 60-sec integration in each exposure on 30 November and on the
first half on 1 December, and a 30-sec integration on the last half on 1 December 
in order to avoid the contamination due to a bright star in FOV. 
The angular velocity of Phaethon's sky motion was rather large in the RA (Right Ascension) direction, 
6--7 arcsec min$^{-1}$, during our observations in Phaethon's close approach to Earth. 
It was comparable to the stellar FWHMs at the observatory,  
and it does not give a big impact to the photometry
under the sidereal tracking for Phaethon. 
As a consequence, we acquired a total of 144 photometric data: 
83 frames of photometric data in total during 4.7 hours on 30 November and 
61 frames of those during 3.5 hours on 1 December, respectively. 
We also measured the flux of several Landolt's standard stars, 
which were imaged under almost the same air mass with Phaethon.
All the photometric data were obtained through the Johnson-Cousins-based $R$-band filter. 
Dark and twilight flat-field frames were taken after the observations in each night.  

All the photometric images were dark-subtracted and flat-fielded  
based on the standard CCD data reduction procedure 
using IRAF, too.  
Then the aperture photometry was done using a circular aperture three times larger than 
the radial profile of Phaethon's FWHM.



\section{Results and Discussion\label{sec:result}}

\subsection{Time-variable spectra of Phaethon\label{ssec:spectralchange}}

Here we deal with the normalized spectral gradient of reflectivity in the VIS-range. 
\citet{luu1990} first reduced the spectral gradients of asteroids, 
fitting the linear-function of wavelength.  
This method is useful for the taxonomic classification of asteroids.  
Therefore, we applied this linear least-squares fit to Phaethon's spectral slope and 
measured its spectral gradient for each frame. 
The normalized spectral gradient of reflectivity, $S'$ (in unit of $\%/0.1$ $\micron$), 
in the wavelength range between $\lambda_1$ and $\lambda_2$ is defined by \citet{luu1990} as  
\begin{equation} 
S'(\lambda_1, \lambda_2) = \left(\frac{dS/d\lambda}{S_{\lambda_{\rm c}}}\right), \label{gradient}
\end{equation}
where $dS/d\lambda$ is a changing rate of the reflectivity gradient 
in the range, $\lambda_1=0.45$ $\micron$ and $\lambda_2=0.65$ $\micron$ in our study. 
$\lambda_{\rm c}$ is the center of this wavelength range, 
so here we take $\lambda_{\rm c} = 0.55$ $\micron$. 
Thus $\lambda_1 \leqslant \lambda_{\rm c} \leqslant \lambda_2$, 
and the spectral slope is represented nearly by the straight line 
in this interval.   
$S'$ serves as a measure of the color of the asteroid, applying the solar color as a standard.  
Hence the color of the asteroid suggests red when $S'>0$, as against blue when $S'<0$. 
The VIS-region contains very limited information on reflectance properties 
for the $S'$-positive taxonomic classes, i.e., S- and C-types.    
However, we can easily discriminate whether or not it is B/F-type class, judging by 
the $S'$-value in the VIS-region, since the B/F-type taxonomic classes only exhibit the negative $S'$-values. 
Hence, here we specially propose calling an extremely blue-slope for Phaethon's VIS-spectrum 
``Phaethon Blue". 
This comes from the VIS-spectrum recorded by \citet{luu1990} on 6 October, 1988 
(the spectrum LJ1988 in Table~\ref{spdata}), and re-measured at $S'=-11$ by \citet{licandro2007}. 
The $S'$-value of Phaethon is more effective and lower in the VIS-range than in the NIR-range,  
by the factor of 10 \citep{licandro2007}.
In fact, the blue-spectral slope of the B-types in the VIS-range 
shows negatively more precipitous than that in the NIR-range 
\citep{Ali-Lagoa2013}.
Therefore, our time-series VIS-spectrophotometry may have more advantage in 
significantly distinguishing spectral flux (or color) variation of Phaethon, if any,  
than applying the NIR-spectroscopy.  

\begin{figure}[htbp]
 \epsscale{0.85}
 \plotone{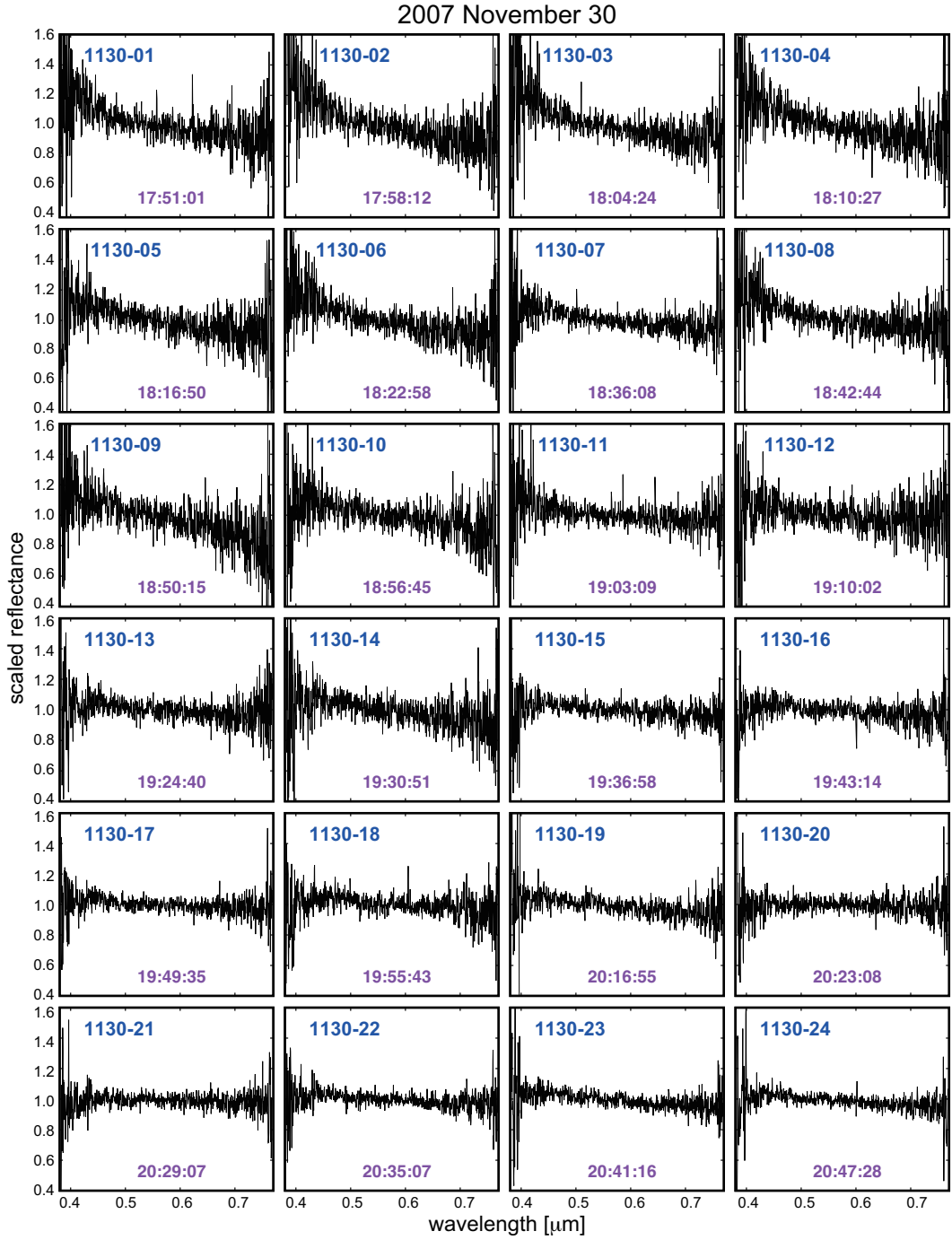}
 \caption{Time-series VIS-spectra of Phaethon recorded on 30 November, 2007.}
 \label{spec_1130}
\end{figure}

\begin{figure}[htbp]
 \epsscale{0.85}
 \plotone{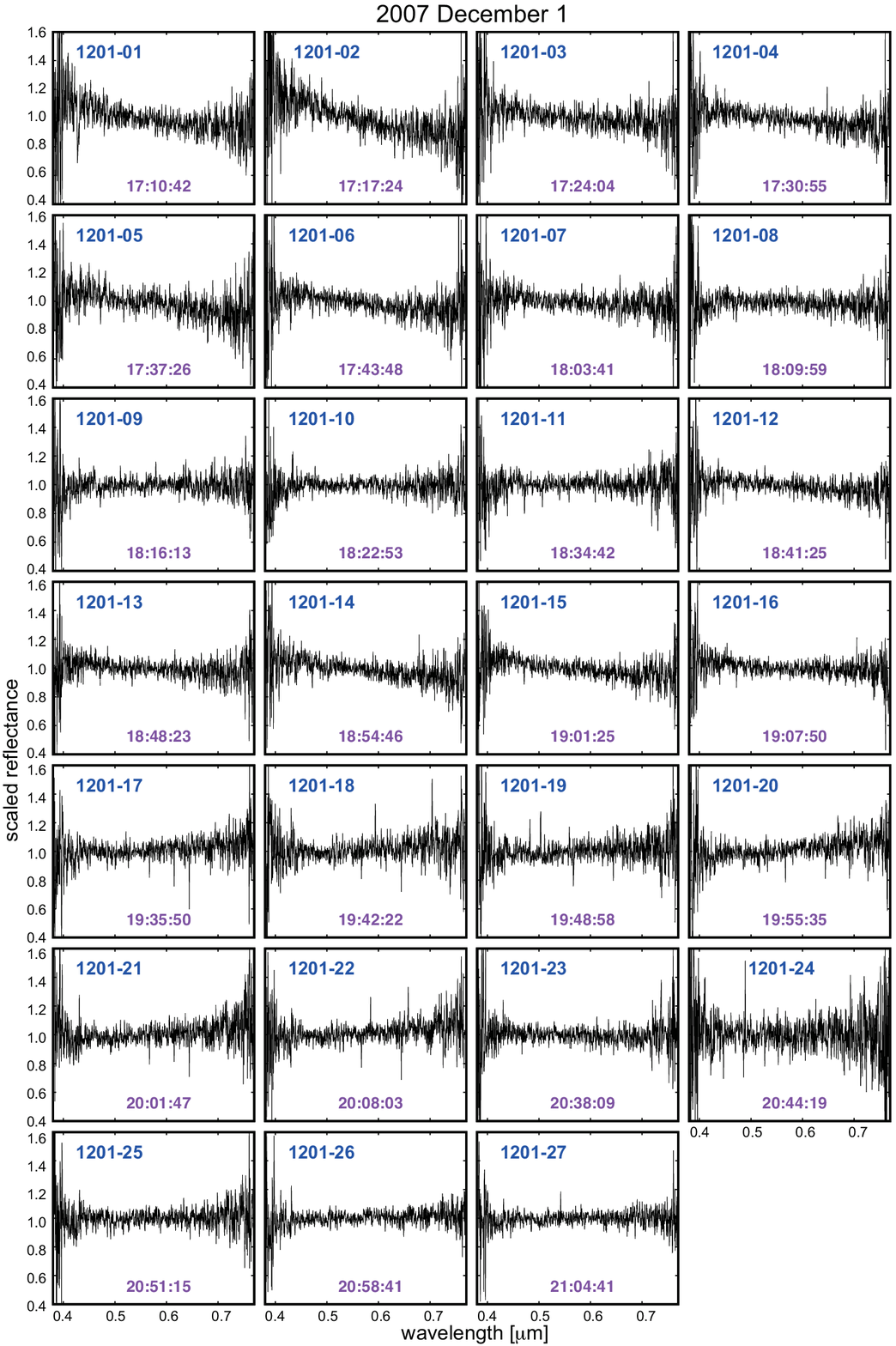}
 \caption{recorded on 1 December, 2007.}
 \label{spec_1201}
\end{figure}

\begin{figure}[htbp]
 \epsscale{0.85}
 \plotone{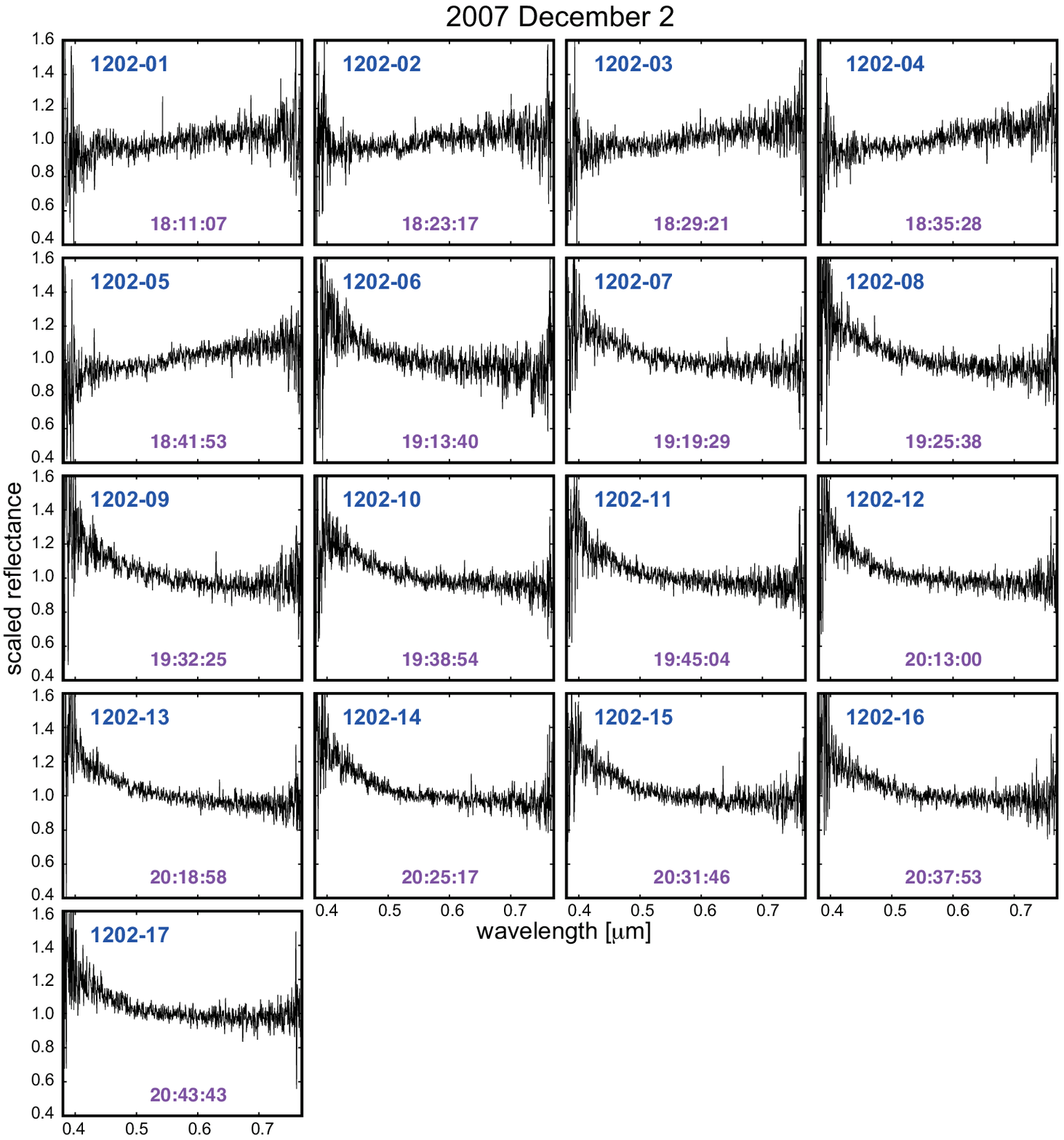}
 \caption{recorded on 2 December, 2007.}
 \label{spec_1202}
\end{figure}

The results of our Phaethon's VIS-spectrophotometries for $S'$ along with their ephemerides and 
geometric parameters are summarized in Table~\ref{obsre},  
The aspect angles (i.e., sub-Earth latitude on Phaethon), 
$\phi_{\rm K}$, $\phi_{\rm A}$, and $\phi_{\rm H}$, are 
calculated on the basis of the pole solution (SOL, hereafter) K by \citet{krugly2002}, 
SOL A by \citet{ansdell2014}, and SOL H by \citet{hanus2016a}, respectively. 
Their pole axis parameters are presented in Table~\ref{psol}.    
Also, the time-series VIS-spectra of Phaethon are patched together in 
Figures~\ref{spec_1130}, \ref{spec_1201}, and \ref{spec_1202} for the 30 November, 
1 and 2 December observations, respectively. 
All the spectra show mostly featureless, as pointed out by several works so far. 
The 0.7-$\micron$ absorption feature \citep[due to Fe$^{\rm 3+}$ abundance in phyllosilicates:][]{vilas1989} 
was not visible in any of our spectroscopic data at all. 
Also, no cometary features were detected in our data, 
although Phaethon sometimes displays the cometary activity near perihelion. 
But, one of our most significant results in this work is that 
Phaethon spectroscopically shows time-variable and 
variety in the taxonomic classification, 
since Phaethon's $S'$ is diverse in the values from 
$-11.45 \pm 1.25$ (so-called, Phaethon Blue) for 
the spectrum 1130-02 (midtime 30 November, 17:58:12 UT) to 
$6.63 \pm 0.26$ for 1202-05 (2 December, 18:41:53 UT). 
It roughly corresponds to the range between B- and C-(or Cb-)type colors  
in the \citet{Bus1999} spectrometric taxonomy, 
suggesting there exists an inhomogeneous surface property on Phaethon, 
as we expected. 
Nevertheless, Phaethon's surface seems to be dominantly covered with the B-type material,   
since a total of 53 out of 68 spectrophotometries, thus $78\%$, are in the negative $S'$-values, 
along with the UV/$B$-band falling-off, 
as against the rest 15 spectrophotometries, $22\%$, being in the positive ones, 
with a weak UV/$B$-band absorption feature, 
as seen in Table~\ref{obsre} and Figures~\ref{spec_1130}--\ref{spec_1202}.
Therefore, the $S'$-values seem to be correlated with the UV/$B$-band feature strengths.  
 
The $S'$-value diversity of our results is almost overlapped by assorted single frame shots 
of the other 15 VIS-spectrophotometries, ranging between $-11$ to $23$,   
as compiled in Table~\ref{spdata}. 
Among those, however the only one red-slope spectrum of Phaethon, CB1983 ($S'=23$),  
has been recorded by \citet{cochran1984}, i.e., $7\%$ of 15 spectra,   
whether stochastically by chance or necessity.
 
Our time-series VIS-spectrophotometry reveals that $S'$ alternately exhibits maximum and minimum with some periodicity,   
gradually changing. 
In fact, the red-slope spectra unevenly cluster at certain time regions, 
as shown in Table~\ref{obsre}: 
only one spectrum (1130-11) on 30 November;  
9 spectra (1201-09, 17, 18, 20, 21, 22, 25, 26, and 27) on 1 December; 
and 5 spectra (1202-01, 02, 03, 04, and 05) on 2 December.  
Therefore, we next carried out periodic analysis for the time-series VIS-spectrophotometries 
based on Phaethon's rotation period   
in order to find whether or not red-colored region is located at the particular rotational phase. 
It may be involved with local (or longitudinal) specific materials and/or structure. 

\subsection{Revisit of Phaethon's rotation period: $P \sim 3.6$ hours or $\sim 7.2$ hours?\label{ssec:period}}

Before examining the rotationally time-resolved spectroscopy,
we revisit the axial rotation state of Phaethon.  
We also analyzed it using our $R$-band photometric data. 
The sidereal rotation period of Phaethon has been reported by several works so far. 
\citet{meech1996} first reduced a ``double-peaked'' rotation
period of $P= 3.604 \pm 0.0011$ hr based on the data acquired on three
nights in December, 1994--January, 1995, from the UH88 telescope of IfA/UH 
(Institute for Astronomy, University of Hawaii) 
and from the Lowell observatory. 
\citet{pravec1998} also obtained two $P \sim 3.6$-hr lightcurves 
on 1 and 2 November, 1997 and in November--December, 2004 
\footnote{the 2004 observations: http://www.asu.cas.cz/~ppravec/newres.txt}, respectively, 
from the Ond\v{r}ejov observatory,   
who derived a synodic ``single-peaked'' lightcurve with $P=3.57 \pm 0.02$ hr based on the 1997 data. 
It is somewhat ``noisy'' along with rather small amplitude, 
which implies that the lightcurve was possibly derived under near pole-on geometry. 
\citet{krugly2002} obtained two $P \sim 3.6$-hr lightcurves with small amplitude, 
using respectively 4-night data in September 1996 and 2-night data in November 1997, 
from the Chuguevskaya station, 
both of which show single-peaked ones, hence similar to that observed by \citet{pravec1998}. 
Then they determined the pole axis orientation, SOL K, as listed in Table~\ref{psol}, 
which is the first candidate between their two SOLs.  
\citet{krugly2007} carried out photometric observations again in 2004, 
when they obtained $P = 3.6052 \pm 0.0008$ hr.
\citet{gaftonyuk2010} obtained $P = 3.604$ hr based on their photometries in 2004, 2006, and 2007 
from the Crimean Astrophysical Observatory. 
\citet{ansdell2014} reduced $P= 3.6032 \pm 0.0008$ hr based on 15-night photometric dataset
in 1994--2014, observed by the IfA/UH team,  
whose lightcurve patterns show a mixture of single- and double-peaked ``looking'' ones 
on different nights, including the data obtained by \citet{meech1996}. 
They also solved the pole axis orientation, SOL A (see Table~\ref{psol}) and 
the lightcurve-inversion convex shape model.    
\citet{warner2015} found another $P \sim 3.6$-hr double-peaked lightcurve from CS3-PDS, 
using 8-night data in November--December, 2014: 
his $P=3.6039 \pm 0.0002$ hr. 
Most recently, \citet{hanus2016a} may have reached the top quality 
in analyzing Phaethon's rotation state, 
who reduced $P=3.603958 \pm 0.000002$ hr, 
using all the available photometric data of Phaethon over the world, 
including the IfA/UH dataset: 
the world's biggest, 49(out of 51)-night dataset in the longest observational arc in 1994--2015. 
Consequently, they performed the reconstruction of fine structure by 
long-term rotational phase curve fitting   
along with the pole axis determination, SOL H (see Table~\ref{psol}) and 
the convex shape model. 
Therefore, number of photometries (or nights) are in order of increasing as SOL K$<$A$<$H. 
   
Hence, the previous works claim $P \sim 3.6$ hr for Phaethon. 
These rotational parameters are compiled in the Asteroid Lightcurve Data Base (LCDB) 
by B.~D.~Warner\footnote{http://www.minorplanet.info/lcdbquery.html}, 
from which we can find that the lightcurve amplitude of Phaethon distributes over 0.07--0.34 mag. 

\begin{figure}[htbp]
 \epsscale{0.7}
 \plotone{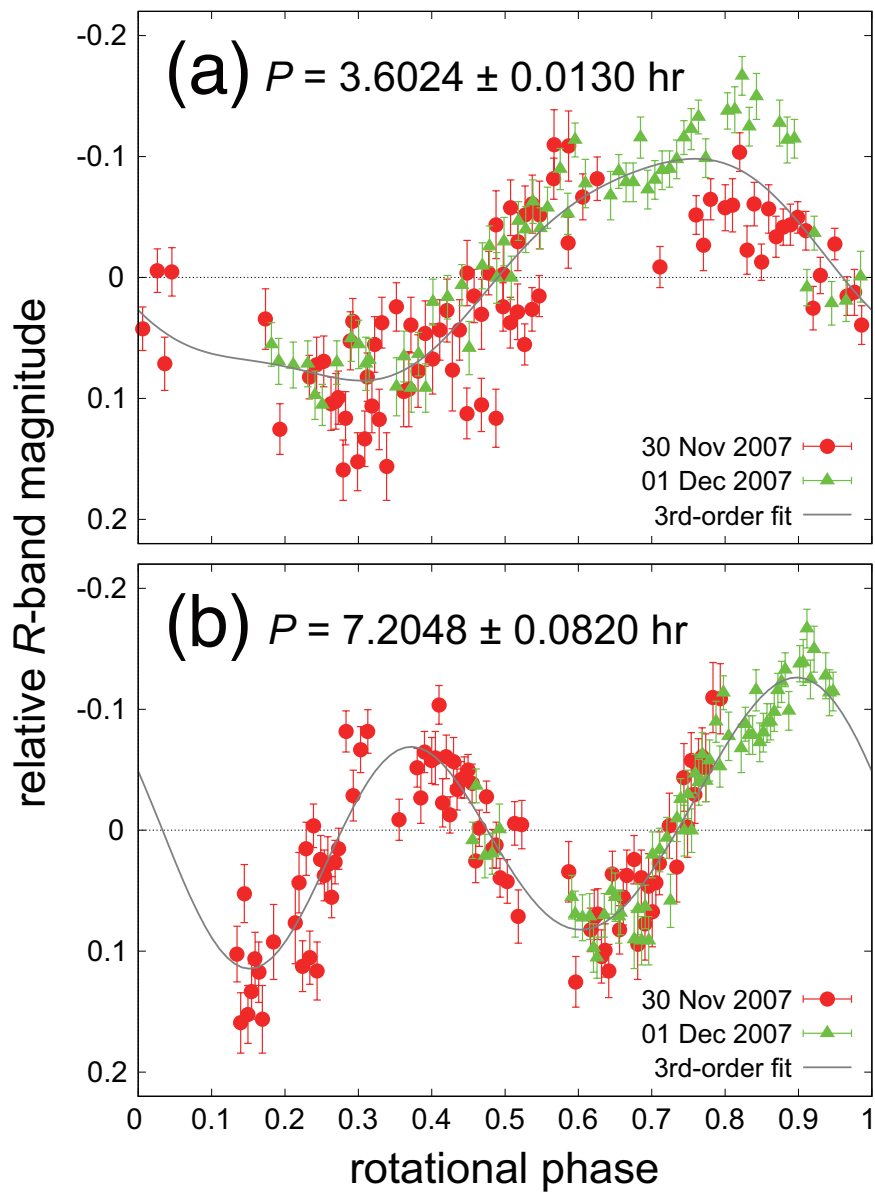}
 \caption{$R$-band photometries and rotational phase-curve fittings of Phaethon. 
 Panel (a) shows the best fitted single-peaked lightcurve  
 by phase-folding with $P = 3.6024 \pm 0.0130$ hr;  
 (b) the double-peaked lightcurve by phase-folding with $2P$, 
 $7.2048 \pm 0.0820$ hr. 
 The gray solid curves are results of the least-square fit 
 by applying the 3rd-order Fourier harmonics. 
 See Section~\ref{ssec:period} for more detail.}
 \label{lc}
\end{figure}

\begin{figure}[htbp]
 \epsscale{0.95}
 \plotone{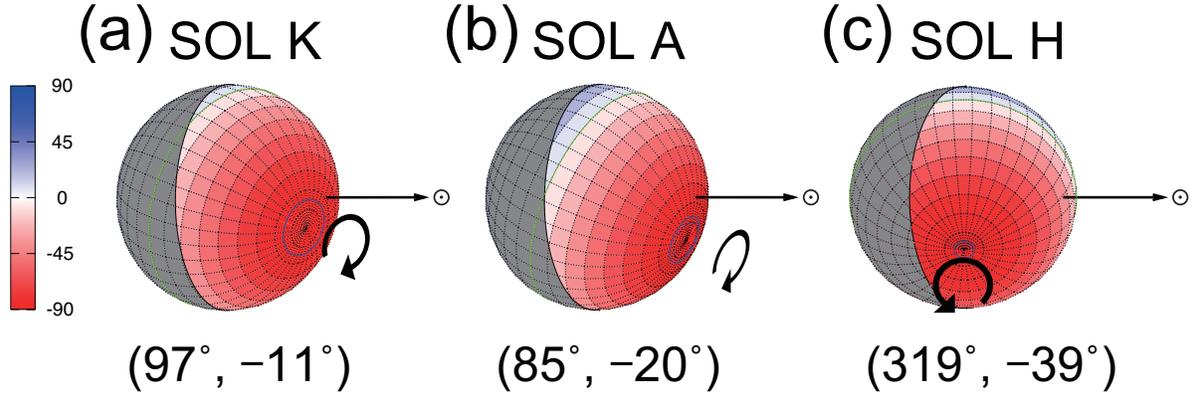}
 \caption{Schematic reconstructions of three geometric models 
 of Phaethon as viewed from the Earth 
 at the time of 19:35:50 UT, 1 December, 2007, 
 at which the spectrum 1201-17 was recorded, 
 i.e., near median time of our spectroscopic observations.  
 Panel (a) is reconstructed from SOL K by \citet{krugly2002} with 
 description of pole coordinate ($\lambda_{\rm P}$, $\beta_{\rm P}$), 
 similarly (b) from SOL A of by \citet{ansdell2014}, 
 and (c) from SOL H by \citet{hanus2016a}. 
 Then their aspect angles are 
 $40^{\circ}.6$S, $26^{\circ}.3$S, and  $61^{\circ}.7$S, respectively. 
 The phase angle, $\alpha$, was $61^{\circ}.90$, 
 thus the ratio of the luminous area to the total area of disk, $k=0.74$, 
 where the gray area indicates the sun-shadow.
 The southern hemisphere faces to the Earth 
 at the time of observation in every models. 
 The arrow directs from the subsolar point to the Sun (symbolized by $\odot$), 
 whose length is equal to Phaethon's radius. 
 The blue line means an error circle for every models.}
 \label{aspect}
\end{figure}

After differential photometry using our $R$-band images, 
we performed the periodicity analysis, 
applying the Period Searching \& Light Curve Fitting Program software ``cyclocode''
\footnote{e.g., available at http://www.toybox.rgr.jp/mp366/lightcurve/cyclocode/cyclocode.html} 
developed by \citet{dermawan2004}, which is an implementation of the Lomb periodogram. 
As a result, we found the sidereal rotation period of Phaethon to be $P = 3.6024 \pm 0.0130$ hr 
based on the best fitted single-peaked lightcurve. 
Also, we got a similar result using the Python module, ``astroML'', for machine learning and data mining for astronomy
\footnote{http://www.astroml.org/}.
It perfectly matches with those previously reduced as above, 
especially the reduction of \citet{hanus2016a} being within the 1$\sigma$-level of our reduction. 
By phase-folding with $P = 3.6024 \pm 0.0130$ hr, 
we obtained the single-peaked synthetic lightcurve, 
as plotted in Figure~\ref{lc}(a). 
The RMS residuals of the fitted lightcurve seem to be somewhat
larger and noisy like the other single-peaked lightcurves obtained 
by \citet{pravec1998} and by \citet{krugly2002}. 
Asteroids as 3D-triaxial ellipsoid bodies usually describe a double-peaked phase curve in one full rotation, 
whereas a single-peaked phase curve may uncommonly occur 
when the body shows bidimensionally near round shape at near pole-on geometry to the Earth  
\citep[e.g.,][]{zappala1980, kaasalainen2005}. 
Therefore, we examined the possibility whether or not sub-Earth latitude on Phaethon was near pole-on situation 
during our observations. 
Three geometric models of Phaethon as viewed from the Earth at the time of 19:35:50 UT, 1 December,  
at which the spectrum 1201-17 was recorded, 
i.e., near median time of our spectroscopic observations,  
are schematically illustrated in Figure~\ref{aspect}(a)--(c), 
assuming the asteroid to be a spherical body.  
They were reconstructed from SOLs K, A and H, 
when their aspect angles were 
$40^{\circ}.6$S, $26^{\circ}.3$S, and  $61^{\circ}.7$S, respectively.    
These aspects are calculated as seen from the geocenter, 
but the parallax of the geocenter with the Lulin observatory is tiny and negligible. 
The phase angle, $\alpha$, was $61^{\circ}.90$, 
thus the ratio of the luminous area to the total area of disk, 
$k=0.5(1+\cos \alpha)=0.74$.
If we define Phaethon's north-pole orientation as the
sunlit side at the perihelion, 
the southern hemisphere faces to the Earth at the time of observation 
in every models. 
Then SOL H exhibits a retrograde rotation against SOLs K and A. 
During the photometric observations, 
$\alpha$ changed a little from $58^{\circ}.87$ to $61^{\circ}.81$, and
likewise $\phi_{\rm K}$, $\phi_{\rm A}$, and $\phi_{\rm H}$ changing  
from $43^{\circ}.5$S to $40^{\circ}.5$S, 
$29^{\circ}.1$S to $26^{\circ}.2$S, 
and $61^{\circ}.9$S to $61^{\circ}.7$S, respectively. 
They do not so strongly affect the RMS residuals of phase curves in our photometries. 
Therefore, Phaethon did not situate in near pole-on viewing geometry in either SOL case 
and its single-peaked lightcurve would be unexpected from our observations. 

On that account, by phase-folding with $2P$, 
we obtained the double-peaked synthetic lightcurve with
the period of $7.2048 \pm 0.0820$ hr, 
as given in Figure~\ref{lc}(b). 
The double-peaked lightcurve visually exhibits a better fitting 
than the single-peaked one, as customary practices, 
although the $1\sigma$-error of $\pm 0.0820$ for this $2P$ phase-folding is 
larger than that of $\pm 0.0130$ for the $P \sim 3.6$-hr phase-folding. 
In addition, photometric data points of the $2P$ phase-folding 
have $\sim 15\%$ coverage gaps in its full rotational phase, 
as another disadvantage.  
Nevertheless, significant differences between the first- and second-cycle patterns and between their amplitudes 
arise in this $2P$ phase-folding, double-peaked lightcurve, 
which denies a possibility of the single-peaked lightcurve solution against Phaethon.  
Hence, these observational facts suggest that $P \sim 7.2$ hr 
is more preferable for Phaethon than $\sim 3.6$ hr along our reduction. 
Then the lower limit of the axis ratio of Phaethon during the photometric observations   
is estimated from the lightcurve amplitude, assuming that the variability
comes entirely from the elongated shape of the object. 
The lightcurve amplitude for $R$-band magnitude, $\Delta m_{R}$, 
is $\sim 0.22$ mag, near the median among LCDB for Phaethon. 
This implies the lower limit of the axis ratio of 1.22 from the relation 
\begin{equation} 
\frac{a}{b} > 10^{0.4 \Delta m_R}, \label{axratio}
\end{equation}
where $a/b$ is the axis ratio of primary and secondary axes. 
This near round shape profile was supported by the Arecibo radar observation 
targeted Phaethon that was operated one-week later of our observations in 2007 
(Lance Benner 2016, personal communication). 
A similar tendency is supported by the convex shape 
derived by \citet{ansdell2014} and \citet{hanus2016a} based on $P \sim 3.6$ hr. 
Both the rotation periods of $P \sim 3.6$ hr and $\sim 7.2$ hr  
do not make any difference in the pole axis determination, 
while they should make a big difference in the shape modeling.  
The $P \sim 3.6$-hr double-peaked-look lightcurve, 
detected by previous works,   
may be caused by Phaethon's partially-angular outline.  

Likewise the sidereal rotation period of 2005 UD based on 
the best fitted single-peaked lightcurve was reduced at 
$P \sim 2.62$ hr, thus double the obtained periodicity, 
$P \sim 5.24$ hr, should be the actual one \citep{jewitt2006, kinoshita2007}.  
In fact, it would be natural that a larger body spins at lower rotation rate 
than smaller fragmentary bodies do 
among identifiable asteroid family members \citep[e.g.,][]{fujiwara1981,kadono2009}. 
Therefore, the rotation periods of Phaethon's $\sim 7.2$ hr and 2005 UD's $\sim 5.2$ hr 
match up well with the size relation between Phaethon and 2005 UD: 
the 5.1 km-size main mass and the 1.3 km-size secondary mass among the PGCs, respectively.

\subsection{Rotational phase variation of the spectral gradients, $S'$\label{ssec:rotvariation}}

\begin{figure}[htbp]
 \epsscale{0.7}
 \plotone{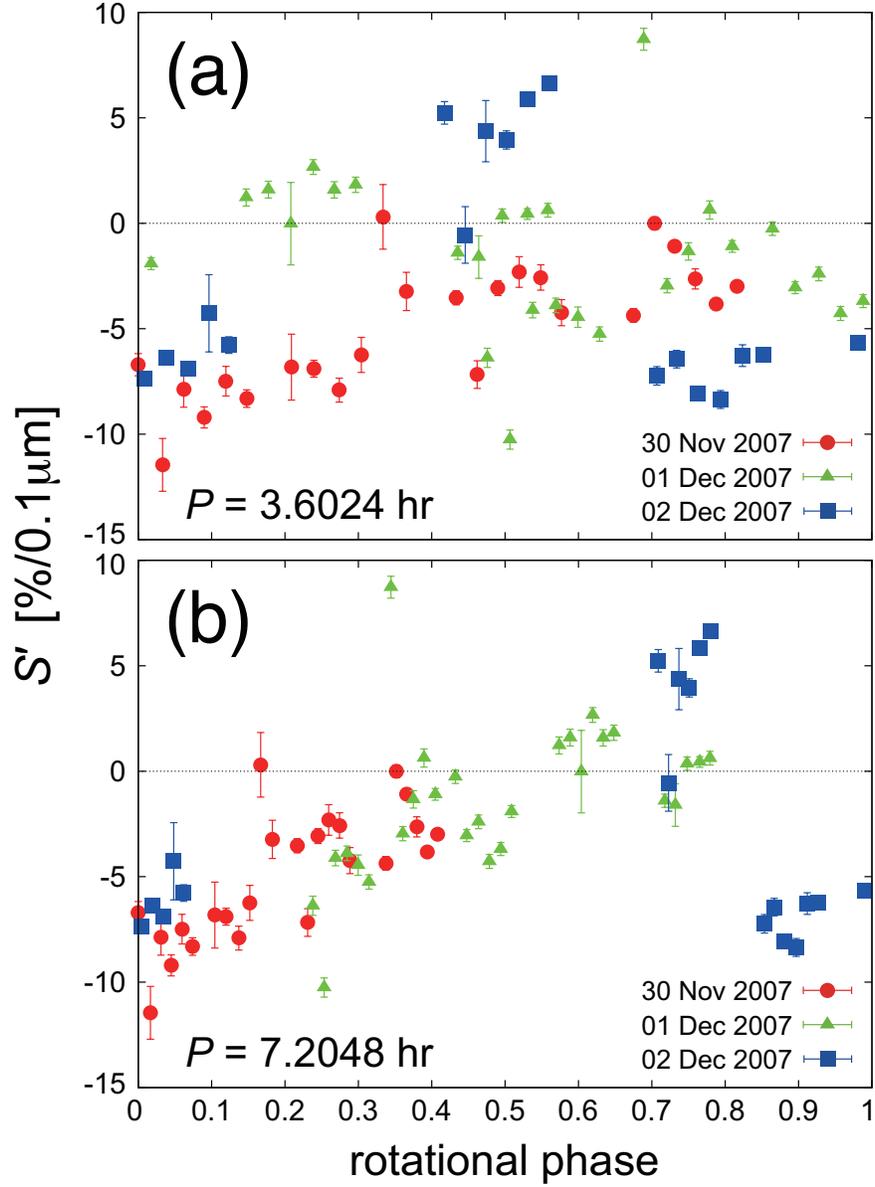}
 \caption{Potential correlation between spectral gradients, $S'$, of Phaethon and
 its rotational phase by phase-folding with $P=3.6024$ hr for panel (a) 
 and $7.2048$ hr for (b).  
 See Section~\ref{ssec:rotvariation} for more detail.   
 }
 \label{sg}
\end{figure}

We finally verify the presence or absence of some correlation between 
spectral gradients of Phaethon and its rotational phase 
by phase-folding with $P=3.6024$ hr and with $7.2048$ hr.  
They are presented in Figure~\ref{sg}(a) and (b), 
where a total of 68 $S'$-data points are plotted on the 
rotational phase by phase-folding with $P=3.6024$ hr for panel (a) and 
with $P=7.2048$ hr for (b). 
When we draw a comparison about them,  
the spectral gradients seem to be almost randomly scattered in panel (a), 
while surprisingly smoother changing in (b).
Panel (a), by phase-folding with $P=3.6024$ hr,  
seems to have a very weak trend of correlation between spectral gradients and rotational phase, 
whose $S'$-data points reach the maximum around the rotational phase $\sim 0.5$ and 
a minimum around $\sim 0$ or $\sim 1$. 
On the other hand, panel (b) by the phase-folding with $P=7.2048$ hr  
obviously reveals rather strong trend of correlation between those.  
The $S'$-data points gradually increase from the rotational phase around 0--0.1 as the minimum, 
and reach the maximum around the phase $\sim 0.7$, after which 
it decreases markedly toward the phase $\sim 1$ of the minimum again.  
Thus this curve shows an asymmetric pattern. 
In addition, it is sure that a possible RMS residual for the $S'$-data points in panel (b) 
seems to be within almost a half of that in (a). 
Also, in any case of $P=3.6024$ hr or $7.2048$ hr, 
the $S'$-values in $\sim 30\%$ of the rotational phase seem to be in the positive zone, 
centered on the rotational phases $\sim 0.5$ and $\sim 0.7$, respectively. 
Hence, we could successfully acquire rotationally-resolved spectroscopic data of Phaethon, 
of which particular rotational phase in fact shows a red-spectral slope as the C-type color, 
as well as 2005 UD, as we expected.
We also ascertain that Phaethon's rotation period of $P=7.2048$ hr is more preferable again,   
in which we observed $\sim 85\%$ of its full rotational phase of the $S'$-data points with 
$\sim 15\%$ coverage gaps.   

Assuming Phaethon's rotation period of $P=7.2048$ hr as correct, 
now we discuss a possible implication of smooth variation of 
the $S'$-values for Phaethon's spectral gradients in Figure~\ref{sg}(b). 
It is natural to infer that the variation of the spectral gradients is caused by a 
longitudinal inhomogeneity rather than a latitudinal one on Phaethon's surface, 
although \citet{ohtsuka2009} expected the latitude-dependent color variations on Phaethon 
based on SOL K of \citet{krugly2002}. 

If the solar-radiation heating is the main determining factor for Phaethon Blue 
and then SOL K is true, 
the solar-radiation heating effects on Phaethon may be a function of the latitude \citep{ohtsuka2009}, 
since Phaethon was then regarded as having a highly-tilted pole axis and thus 
its mean precession rate can be negligible. 
These make Phaethon's pole orientation to be 
almost fixed and stable against the ecliptic plane for the long term. 
Moreover, Phaethon's orbital motion evolves slowly
under the secular perturbation, mainly driven by the Lidov--Kozai circulation \citep{urakawa2014}. 
Thus, the northern hemisphere would be selectively heated up to 
the temperature at 800--1100 K level near perihelion.  
Contrary, then the southern hemisphere may be 
heated to lower temperature at 600--800 K at most, 
thus barely exceeding 
the serpentine-phyllosilicate decomposition and dehydration temperatures at 
573 K and 673 K, respectively. 
If so, the southern hemisphere may suffer a lesser degree of metamorphism  
than the northern hemisphere. 
However, we cannot find such a tendency from other spectrophotometric works 
(in Table~\ref{spdata})
along with our works. 
In addition, we have recorded Phaethon Blue, $S'=-11.45 \pm 1.25$ for the spectrum 1130-02, 
despite the southern hemisphere dominantly facing to the Earth (then $\phi_{\rm K} = 43^{\circ}.21$S), 
which is bluer than any other spectrophotometry for Phaethon and B/F-type asteroid, 
and comparable with $S'=-11$ for the spectrum LJ1988 \citep{luu1990}, 
previously the bluest record for Phaethon.  
Then LJ1988 was observed when the southern hemisphere facing to the Earth again, 
and when $\phi_{\rm K}=53^{\circ}.8$S.
The spectrophotometries for CB1983, Ch1993, and Li2003 (see Table~\ref{spdata}), 
recorded when facing the northern hemisphere to the Earth, 
obviously manifest that the northern hemispheric surface of Phaethon is not always 
bluer than Phaethon Blue.    
This discrepancy will be little change if we apply SOL A of \citet{ansdell2014}, 
since the spin axes of SOLs K and A trend in almost the same direction each other. 
It should be, however, noted that 
\citet{ansdell2014} found a possible latitude-dependent color variation on Phaethon 
from their multi-color photometric data.

On the other hand, if we take SOL H, 
things are largely different from those taking SOLs K and A. 
In fact, \citet{hanus2016a} speculated that SOLs K and A were incorrectly 
determined, since Phaethon's surface color do not exhibits a latitude-dependent variation 
\citep{ohtsuka2009}.  
From this background, they reduced another preferential pole solution, SOL H,    
using larger photometric dataset in longer photometric arc 
than SOLs K and A, mentioned above. 
In fact, its error circle is the least estimated at $\pm 5^\circ$ 
among three SOLs, and we may regard SOL H as the most reliable one. 
\citet{hanus2016a} demonstrated that 
their pole orientation should be highly changed by secular pole axis precession
and the southern hemisphere is also preferentially heated by solar radiation heating 
around perihelion at present evolutionary epoch.  
If so, the entire surface, including the southern hemisphere, 
has long been almost equally burnt out by the solar radiation heating effects 
until now, which may saturate the thermal metamorphism in Phaethon's entire surface, 
and also equally space-weathered, making Phaethon Blue.
Therefore, there are no latitude-dependent variations on Phaethon, according to SOL H.
If it is true, there never exists primitive chondritic materials 
on Phaethon's entire surface, except for an exposition of likely primordial internal structure
due to a large-scale splitting or breakup event long ago on Phaethon (or PGC precursor). 

The aspect angles, $\phi_{\rm H}=61^{\circ}.02$S--$61^{\circ}.86$S, in our observations, 
have ever been in fact at the highest range in Phaethon's southern hemisphere 
among all the VIS-spectroscopic observations, which maybe a special case. 
The spectrum La1994 recorded $S'=-3.9$ under the second-highest aspect, 
$\phi_{\rm H}=40^{\circ}.3$S \citep{lazzarin1996}, 
as listed in Table~\ref{spdata}, 
however normal color as Phaethon.
Note that the only one red-slope spectrum, CB1983 with $S'=23$,  
was obtained under $\phi_{\rm H}=16^{\circ}.3$N., 
the second-highest northern-hemispheric latitude, 
maybe another special case. 

\subsection{Color composition on Phaethon's surface with C-type red-colored area \label{ssec:colcomp}}
\begin{figure}[htbp]
 \epsscale{0.3}
 \plotone{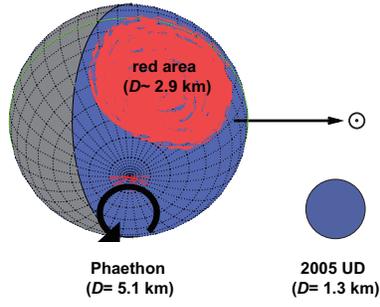}
 \caption{Conceptual scheme of the C-type red-colored area on Phaethon 
 as viewed from the Earth at the time of which 
 the spectrum 1202-01 ($S'= 5.23 \pm 0.54$) was recorded. 
 The pole orientation is based on SOL H \citep{hanus2016a}.
 See Section~\ref{ssec:colcomp} for more detail. 
 }
 \label{ra}
\end{figure}

As we have seen, we detected only two spectra with $S'<-10$: 
1130-02 (Phaethon Blue) and 1201-02 with $S'=-10.26$. 
Next, we consider the color composition in Phaethon's surface. 
Let us imagine that there are two distinct 
areas with two extreme pure colors on Phaethon's surface, 
i.e, Phaethon Blue and the C-type red color.  
Here we define two spectral gradients for their original colors 
as $S_{\rm B}'$ and $S_{\rm R}'$, respectively.  
Then we can presume the observed (apparent) spectral gradient, $S_{\rm O}'$, 
composed by the mixture of  both colors with the aria ratio 
$(A_1:A_2)$ between $S_{\rm B}'$ (for $A_1$) and $S_{\rm R}'$ (for $A_2$), as    
\begin{equation}
 S_{\rm O}' \sim f(\alpha) \left(\frac{A_1 S_{\rm B}' + A_2 S_{\rm R}'}{A_1+A_2}\right), \label{aratio}
\end{equation} 
where $f(\alpha)$ is the phase reddening function, 
$f(\alpha) \sim 1$ when $\alpha >40^{\circ}$ \citep{luu1990},  
although $f(\alpha)$ for the C-complex asteroids has been not well-studied yet.
Assuming $S_{\rm B}'$ and $S_{\rm R}'$ to be two extreme color-slopes for Phaethon  
ever recorded, $-11$ and 23, 
we can reconstruct $S_{\rm O}' \sim 5$, 
i.e., the $S'$-maximum among our spectrophotometries, 
such as at the rotational phase around $0.7$ in Figure~\ref{sg}(b).
Consequently, the red-colored area needs at least involving $\sim 45\%$ of Phaethon's luminous area, 
which results in at least $33\%$ of Phaethon's disk,  
suggesting the red area to be $\sim 2.9$ km in diameter.  
This is considerably larger than 2005 UD's cross-section area (of 1.3 km in diameter), 
as illustrated in Figure~\ref{ra}. 
If we assume $S_{\rm R}' \sim 5$, the red area needs $100\%$ of Phaethon's luminous area 
and thus at least $74\%$ of Phaethon's disk. 
Therefore, the larger $S_{\rm R}'$ is, the smaller the red area becomes, according to Equation~(\ref{aratio}). 
In any case, there exists a certain large C-type red-colored area 
somewhere, maybe mid-latitude in Phaethon's southern hemisphere, 
while Phaethon's surface would be dominantly covered 
by the B-type-colored (or Phaethon Blue) material, 
suggesting thermally metamorphosed CI/CM chondrites.   
A part of the red area might be always face-on to the Earth during our VIS-spectroscopic observations, 
possibly including the SP region. 
The asymmetric $S'$-variation pattern, shown in Figure~\ref{sg}(b), 
would be due to disappearance of the red area under the sun-shadow
and/or due to a non-circular asymmetric shape of the red area.

The petrogenesis for the C-type red-colored area may be due to a impact-excavation 
on Phaethon (or PGC precursor) with the splitting or breakup event 
that happened long time ago,  
when a large amount of PGCs would be released to the cosmic space, 
exceeding the mass of 2005 UD. 
Or contraction cracking due to repeated solar radiation heating $>$1000 K near perihelion 
may have triggered a splitting or breakup event on Phaethon \citep{granvik2016}. 
Therefore, Phaethon's surface and/or subsurface should be easily decomposed due to 
high-temperature heating \citep{akai1992,delbo2014}, 
e.g. shock and solar-radiation heating effects, 
which may produce larger C-type area than we supposed. 
CI and CM chondrites are generally more friable and crumble than any other meteorite type, 
including CK chondrites \citep[e.g.,][]{tomeoka2003}. 
In addition, it is likely that 
regolith grains are released from the C-type red-colored area as cometary dusts: 
matrix-phyllosilicate grains of CI/CM chondrites are in $\micron$-order in size, 
thus well matching with the $\beta$-syndyne dust particles released from Phaethon.  
On the other hand, regolith grains in the Phaethon Blue area would aggregate 
by partially ``unglazed''-like sintering effect with increasing heating time
at subsolidus temperature of silicates; thus growing from fine to coarse grains. 
However, if we take $S_{\rm R}' \sim 23$, then the red area will be fairly red, 
and thus may not be of the C-types:  
\citet{cochran1984} has indeed classified Phaethon as S-type 
on the basis of its relatively high $S'$-value.
Therefore, our another scenario is that the red area consists of  
an ejecta blanket of the S-type impactor:  
in that case NEA 2011 XA$_{3}$ may be a remnant \citep{urakawa2014}.
If $S_{\rm R}' \sim 5$, we may regard it to be still in the C-type taxonomy. 

If we assume that the center of the red-colored area is located on the latitude of $45^{\circ}$S, 
as shown by Figure~\ref{ra},
we realize on the basis of SOL H that 
this area maximally absorbs heat energy from the Sun 
at $1.03$ day before the perihelion passage of Phaethon. 
It is interesting that this is almost consistent with the dust-tail forming timescale of $\sim 1$ day 
just before the perihelion \citep{jewitt2013b,hui2017}.  
Then the local equilibrium surface temperature at $45^{\circ}$S will be estimated 
to reach near maximum at 740--1000 K based on the FRM and NEATM, 
suggesting the temperature high enough for saponite and serpentine decomposition and dehydration.  
Hence, this may trigger thermal fracture in the red area, 
from which $\sim 1$ $\micron$-order radius dusts release.       
In addition, the surface materials even in the red area would never be 
a water-reservoir due to periodic solar radiation heating. 
In fact, no 0.7-$\micron$ absorption feature was visible 
in any of our spectroscopic data, 
probably due to thermal-decomposition of Fe-bearing phyllosilicates
as mentioned above. 
The $S'$-values of all the other spectrophotometries, listed in Table~\ref{spdata}, 
may also be reconstructed by the mixture of the Phaethon Blue and the red-colored areas. 
A redness degree of the red-colored area would change into neutral--blue with time in the future, 
if the area is not an ejecta blanket of the S-type impactor.  

\subsection{Meteoritic analog for Phaethon \label{ssec:phblue}}
Here we discuss on the meteoritic analog and 
the petrogenesis of Phaethon's surface blue-color (or Phaethon Blue).  
\citet{ohtsuka2009} proposed that CI/CM chondrites heated at high- temperature 
may be the candidate analog for Phaethon's surface.
Recently, CK chondrites are suggested as another meteoritic analog candidate 
for Phaethon \citep{clark2010,hanus2016b}.  
CKs are also thermally metamorphosed at high-temperature ranging 550--1270 K \citep{chaumard2012}, 
which have undergone internal heating 
since CKs are supposed to come from the area near central core of the CV-CK clan modeled 
parent body \citep{greenwood2010}.
In terms of the asteroid taxonomy, 
CV/CK chondrites would be originated from 
the K-type Eos collisional family \citep{mothe-diniz2008,greenwood2010}. 
CK chondrites generally exhibit a blue-slope in the VIS-NIR range,
whereas CV chondrites, located on more exterior--surface part of the modeled parent body, 
exhibit a red-slope \citep{cloutis2012b,cloutis2012c}. 
Hence, it is possible that Phaethon's surface is covered by 
the CV/CK-related materials. 
Space-weathering effects on CK-chondritic bodies have never been known yet. 

For searching Phaethon's meteoritic analog, 
we focus on the UV/$B$-band absorption feature 
which is very weakly seen in our VIS-spectrophotometric data. 
This feature presumably comes from the decomposition of Fe-bearing phyllosilicates 
in CI, CM, and CR chondrites as well as the 0.7-$\micron$ feature \citep{vilas1989}.  
Generally, the UV/$B$-band falling-off works in conjunction with 
the less remarkable 0.7- and 3-$\micron$ features 
with increasing the heating temperature \citep{hiroi1996}. 
So, we measured the spectral gradients, $S'$, 
in the VIS-spectra of metamorphic carbonaceous chondrite samples  
through Equation~(\ref{gradient}): 
e.g. CI/CM anomalous chondrites underwent high-temperature heating, 
experimentally heated CI/CM chondrites, and CK chondrites, 
in various grain-sizes. 
Also, we picked up several CM chondrites for reference 
to confirm the grain size effect. 
These data are available by the Keck/NASA RELAB multi-user spectroscopy facility
\footnote{http://www.planetary.brown.edu/relab/}. 
The results of the $S'$-measurements are presented in Table~\ref{relab}, 
where we can see that only three spectra seem blue:  
Yamato (Y-)82162 (CI-an) chip is the only one natural meteorite 
that has a blue-slope with $S'=-0.32$. 
\citet{nakamura2005} classified this metamorphic chondrite 
at his heating stage $\rm I\!I\!I$ (heated at $<750^\circ$C). 
The experimentally-heated Ivuna (CI1) at $700^\circ$C in the grain size $<125$ $\micron$  
has the bluest-color among all, $S'=-4.24$. 
The experimentally-heated Murchison (CM2) at $900^\circ$C in 63--125 $\micron$ also 
shows a blue-slope with $S'=-0.98$,  
as against the meteorite in $<63$ $\micron$ showing a red-slope with $S'=0.26$.  
The principal difference between those $S'$-values is probably due to the grain size effect. 
Belgica (B-)7904 and Y-86720 are CM anomalous chondrites that experienced intense heating and dehydration, 
both of which are categorized at Nakamura's heating stage $\rm I\!V$ ($>700^\circ$C). 
However, they never exhibit a blue-slope in the VIS-range. 
It is interesting that PCA02012 (CM2) exhibits the reddest-slope, $S' =28.85$, 
almost comparable to Phaethon's spectrum, CB1983 ($S'=23$), 
which may also indicate a C-type red-color, rather than S-type. 
No CK chondrite has a negative $S'$-value ($S'=0.34$--$7.29$) 
due to relatively strong UV/$B$-band absolution feature,  
since the CK-matrices never contain phyllosilicates. 
In addition, CK chondrites in fact show a strong silicate absorption feature 
of olivine at around 1.05 $\micron$, 
because well-crystalline olivine is a major phase of CKs 
\citep{cloutis2012c}. 
But, either Phaethon and metamorphosed CI/CM chondrites do not possess  
this feature \citep{licandro2007}. 
The $S'$-values for the RELAB CV and CO chondrite samples range in  
3--8 and 10--20, respectively. 
The grain size effects are obviously recognized on all samples, 
as shown in Table~\ref{relab}.  
We summarized $S'$ and several absorption features of Phaethon and of metamorphic chondrites 
as meteoritic analog candidates for Phaethon in Table~\ref{analog}. 
Based on matching the absorption features between them, we speculate 
thermally metamorphosed CI/CM chondrites as likely candidate meteoritic analogs for Phaethon. 
Especially, the heated Ivuna at $700^\circ$C in $<125$ $\micron$ deserves to be 
the top candidate analog among all the RELAB samples. 

Nevertheless, there exists considerable gap in the $S'$-values between   
Phaethon Blue and metamorphosed CI/CM samples.
$S' \sim -11$ for Phaethon Blue has not been reproduced 
by the RELAB spectroscopic experiments yet.
Phaethon would be the most metamorphic C-complex asteroid in our solar system, 
along with heavily space-weathered in the near-Sun environment.  
Consequently, Phaethon Blue became the most extreme blue among the SSSBs. 
Having Pallas' surface and/or subsurface color as the initial condition, 
Phaethon Blue may be created through additive color processes due to 
several combined effects, such as:  
\begin{itemize}
\item Short-duration (shock) heating effect---Phaethon may be split and broken up by an impact event, 
around which surface materials would be highly metamorphosed.  
\item Long-duration heating effect---Phaethon should be recurrently heated at $\sim 1000$ K around its perihelion. 
\item Grain size effects---Phaethon's surface may be covered by regolith grains larger than 100 $\micron$-order size. 
\item Space-weathering---the fluxes of solar wind and micrometeorite bombardments increases as the asteroid approaches the Sun 
\citep{mann2004}.
\end{itemize}   
Those processes should substantially change Phaethon's surface color, 
darker and bluer with time.  
Hence, we consider that Phaethon Blue may be the saturated color not only by thermal metamorphism 
but also space-weathering 
along with the grain size effects due to regolith grains $>$100 $\micron$. 
We may be able to replicate Phaethon Blue using CI chondrites by laboratory experiments.  

Note that the short-duration shock heating under high-pressure 
may induce a incipient partial melting process.  
This will make achondritic materials 
around the impact area, if PGCs were of impact origin \citep{arai2012}. 

\subsection{Future opportunity to observe Phaethon: in 2017 \label{ssec:fopp}}
\begin{figure}[htbp]
 \epsscale{0.7}
 \plotone{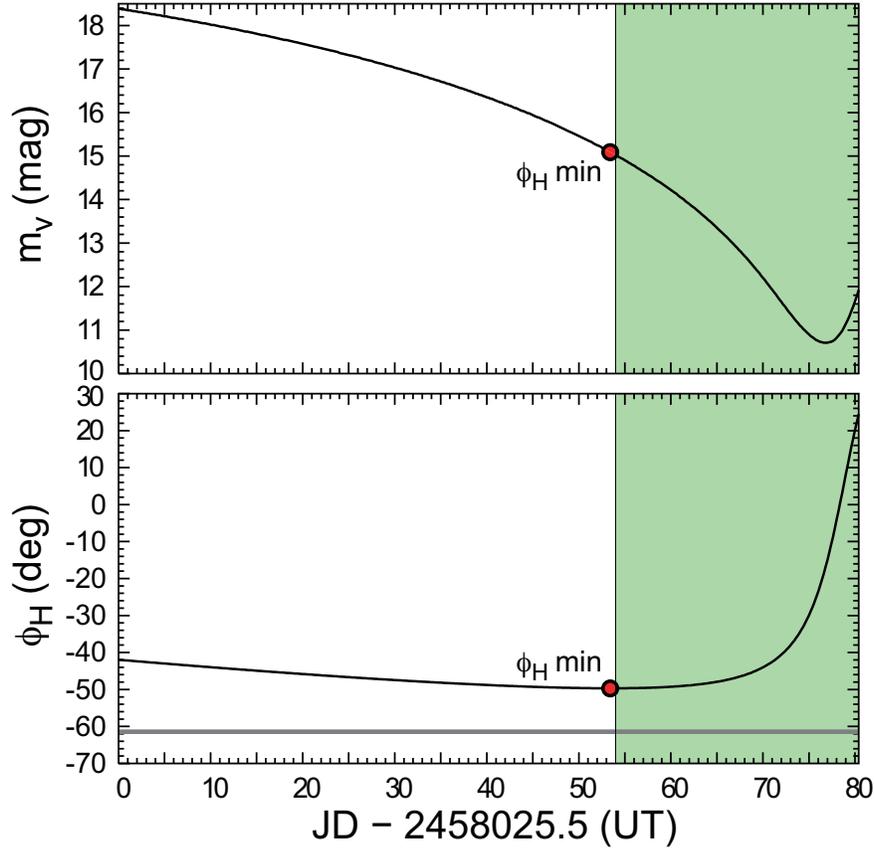}
 \caption{The time-variation of $\phi_{\rm H}$ (lower panel) along with $m_{v}$ (upper panel) in 2017.  
 The abscissa is time in Julian Date (JD) since 29 September, 2017 (JD 2458025.5) 
 up to 18 December, 2017 (JD 2458105.5), 
 in which the observing conditions at a solar elongation of over $90^{\circ}$ turn up. 
 The horizontal gray lines between $61^{\circ}.02$S to $61^{\circ}.86$S indicate 
 the geometric conditions of our spectrophotometries in 2007. 
 The light-green zone in UT indicates the brightness of Phaethon to be $m_{v}<15.0$ mag  
 in 22 November--18 December,   
 thus providing good observing conditions for spectrophotometry using the 2-m class telescopes.   
 The aspect $\phi_{\rm H}$ will reach down to minimum (=highest) at $49^{\circ}.7$S on 21 November  
 (marked by the red circle), 
 in which the difference of $\phi_{\rm H}$ with our VIS-spectroscopic observations in 2007 is only $\sim 12^{\circ}$. 
 So, Phaethon will be observable in several days around 21 November, 2017 under similar geometric condition with 
 the 2007 observations for testing our hypothesis.
 } 
 \label{phh_eph}
\end{figure}

As discussed above for our results, several open questions
still remain concerning Phaethon's nature. 
Further ground-based observations will give us some hints for these questions.
Especially, we have a good opportunity to confirm those 
during Phaethon's next close approach to the Earth in 2017. 
Then Phaethon will flyby the Earth by minimum geocentric distance of 0.0689 AU with 
relative velocity of 31.9 km sec$^{-1}$ on 16 December, 
around which Phaethon will be brighten up to $m_{v} \sim 10.7$ mag. 
Here we estimated the ephemerides of $\phi_{\rm H}$ in 2017, 
whose time-variation is plotted in Figure~\ref{phh_eph},  
in which the observing conditions at solar elongation of 
over $90^{\circ}$ turn up. 
Phaethon will be $m_{v}<15.0$ mag in 22 November--18 December,  
thus providing good observing conditions for Phaethon's spectrophotometry using the 2-m class telescopes. 
The minimum (=highest aspect) $\phi_{\rm H}$ will reach down to $49^{\circ}.7$S on 21 November  
(marked by the red circle in Figure~\ref{phh_eph}), 
in which the difference of $\phi_{\rm H}$ with our VIS-spectroscopic observations in 2007 is only $\sim 12^{\circ}$. 
Incidentally, then $\phi_{\rm K}$ and $\phi_{\rm A}$ will be $64^{\circ}.4$S and $51^{\circ}.7$S, respectively, 
thus also locating near mid--high latitude in the southern hemisphere.
So, Phaethon will be observable in several days around 21 November, 2017 under similar geometric condition with 
the 2007 observations,  
when our hypothesis that there exists the large c-type red-colored  area in Phaethon's southern hemisphere 
could be tested by another VIS-spectroscopy. 
$\phi_{\rm H}$ will rapidly change from the southern to northern hemisphere around Phaethon's close approach to the Earth.  
Therefore, it is important for constraining the size of the red area 
to measure Phaethon's $S'$ by the VIS-spectroscopy   
under different aspect angles.. 

Furthermore, $in$-$situ$ observations of Phaethon by a space
mission should be by far the best way to elucidate  what we have questioned. 
Several scientific mission proposals to Phaethon have been suggested so far 
\citep[e.g.,][]{belton1999,kasuga2006}.
In fact, now Phaethon is selected as one of the top priority candidates for 
the JAXA DESTINY+ 
(Demonstration and Experiment of Space Technology for INterplanetary voYage) mission target object 
\footnote{https://www.lpl.arizona.edu/jaxaworkshop/}. 
The space mission that dispatches the DESTINY+ spacecraft and makes it flyby Phaethon near its descending node 
with relative velocity of 33--34 km sec$^{-1}$ 
has already been designed by \citet{sarli2015}. 
This spacecraft is planned to have an onboard telescopic and multi-band cameras and also a dust analyzer. 
If the DESTINY+ mission is successfully achieved, 
it will make almost all of our questions on Phaethon answered. 
It will also serve as a complement of the OSIRIS-REx sample return mission that 
targets another B-type NEA (101955) Bennu \citep{lauretta2015} 
\footnote {see also http://www.asteroidmission.org/}. 
However, we should sufficiently acquire physical, geological, mineralogical,  and
dynamical information on Phaethon from further ground-based observations 
prior to the DESTINY+ mission.   

\section{Summary and Conclusion\label{sec:conclusion}}

We hypothesize that if the PGCs were generated by
any splitting or breakup event on Phaethon (or PGC precursor), 
the C-type-like surface,
implying the primordial internal structure of the precursor, may still
expose somewhere in Phaethon's surface after impact excavation.
If it is true, some rotational part of Phaethon should exhibit the red-spectral slope as the C-type color 
as well as another large PGC candidate, 2005 UD.
To ensure this working hypothesis, we carried out the time-series VIS-spectroscopic
observations of Phaethon in 30 November--2 December, 2007, 
along with the $R$-band photometric observations, 
covering more than one full rotational phase, 
during which we recorded 68 VIS-spectra and 144 $R$-band photometric data.  
All the spectra show mostly featureless. 
But they show obviously time-variable and variety in the taxonomic classification, 
of which spectral gradients diverse between $-11.45 \pm 1.25$ (so-called, Phaethon Blue, 
which is bluer than any other SSSB)
and $6.63 \pm 0.26$, thus ranging between B- and C-(or Cb-)type colors---
an inhomogeneous surface property on Phaethon as expected. 
The $S'$-value diversity of our results is almost overlapped by assorted single frame shots 
of the (past) other VIS-spectrophotometries, ranging between $-11$ to $23$. 

To find whether or not red-colored region is located at the particular rotational phase, 
we next carried out periodic analysis for our time-series spectra, 
applying our reduced rotation periods of Phaethon.   
Based on our photometries, 
a single-peaked sidereal rotation period of Phaethon was found to be 
$P\sim 3.6$ hr as pointed out by previous works: our $P = 3.6024 \pm 0.0130$ hr. 
However, we reduced another rotation period, $P = 7.2048 \pm 0.0820$ hr, 
by the double-peaked lightcurve fitting, 
which seems more preferable solution.  
The reason is then we did not observe Phaethon under near pole-on situation 
if according to either of the pole solutions, SOLs K, A, and H.  

We verify the presence or absence of some correlation between 
spectral gradients of Phaethon and its rotational phase 
by phase-folding with $P=3.6024$ hr and $7.2048$ hr. 
We found a very weak trend of correlation between spectral gradients and rotational phase 
by phase-folding with $P=3.6024$ hr, as against revealing rather strong trend of correlation 
by phase-folding with $P=7.2048$ hr.  
Hence, here we ascertain again that $P=7.2048$ hr is more preferable rotation period for Phaethon. 
In any of these cases, we successfully acquire rotationally time-resolved spectroscopic data of Phaethon, 
of which particular rotational phase shows a red-spectral slope as the C-type color, 
as 2005 UD dose, suggesting longitudinal inhomogeneity on Phaethon's surface. 

If the solar-radiation heating is the main determining factor for Phaethon Blue 
and then SOL H that we regard as the most reliable pole solution is true, 
Phaethon's entire surface has long been almost equally burnt out by solar radiation heating
until now.
This may result in a saturation of the thermal metamorphism in Phaethon's entire surface, 
and also equally space-weathered, causing Phaethon Blue. 
However, we detected only two spectra having $S'<-10$ among all, including one Phaethon Blue. 
Therefore, we next consider the color composition for Phaethon's surface color. 
If we assume Phaethon's surface is simply covered by two distinct 
areas with two extreme pure colors, i.e, Phaethon Blue and the C-type red color, 
we can presume and reconstruct the observed (or apparent) spectral gradient. 
As a result, we conjecture that the C-type red-colored area is considerably larger than 
2005 UD's cross-section area ($\sim 1.3$ km), suggesting the PGCs  
have an impact-induced origin. 

We also surveyed the meteoritic analog of Phaethon's surface blue-color  
and found thermally metamorphosed CI/CM chondrites to be likely candidates. 
However, there exists considerable gap between the $S'$-values of  
Phaethon Blue and metamorphosed CI/CM samples. 
Hence, we consider that Phaethon Blue may be the saturated color not only by thermal metamorphism 
but also space-weathering along with the grain size effects due to regolith grains $>$100 $\micron$.

We will have a good opportunity to confirm our VIS-spectrophotometric results  
during Phaethon's next close approach to the Earth in 2017. 
Phaethon will be observable in several days around 21 November, 2017 
under the similar aspect condition with 
our VIS-spectroscopic observations in 2007. 
At this opportunity,  
our hypothesis can be tested through the spectroscopy, 
and we will eventually know whether or not there is the large C-type red-colored area in Phaethon's southern hemisphere. 
   

\section*{Acknowledgement}


A major part of our observations was carried out at Lulin Observatory
operated by National Central University (NCU), Taiwan.
The authors are grateful to the local supporters at Lulin:
Jing-Chuan Du, Hao-Wei Shih, Jun-Shiung Shih, and Zong-Jing Wan.
A part of the data analysis was carried out at Center for Computational
Astrophysics (CfCA), National Astronomical Observatory of Japan (NAOJ).
Several spectroscopic data of Phaethon utilized here were obtained and made
available by the MIT--UH--IRTF Joint Campaign for NEO Reconnaissance.
Takahiro Hiroi gave us useful information on the spectroscopic data of
carbonaceous chondrites,
available by the Keck/NASA RELAB multi-user spectroscopy
facility at Brown University.
Alessondra Springmann introduced Lance Benner to us, who provided us with
critical information on his unpublished radar observation in 2007
that gave us important clues for estimating the shape of Phaethon.
The authors have benefited from stimulating enlightenment through 
discussions
with Tomohiko Sekiguchi, Masateru Ishiguro, Fumi Yoshida, and Sunao 
Hasegawa.
DK thanks Center for Planetary Science (CPS), Kobe University,
for providing us with opportunities of interaction and discussion with
number of planetary scientists.
This work is supported by
the research grant from National Science Council of Taiwan (ID 99--2112--M--008--013),
the Ministry of Education of Taiwan under ``Aim for Top University Program,''
and
the Kakenhi Grant of Japan Society for the Promotion of Science (JSPS),
JP25400458/2013--2016 and JP16K05546/2016--2018.


\begin{thebibliography}{}
\expandafter\ifx\csname natexlab\endcsname\relax\def\natexlab#1{#1}\fi

\bibitem[{Akai(1992)}]{akai1992}
Akai, J. 1992, Sixteenth Symposium on Antarctic Meteorites. Proceedings of the
  NIPR Symposium, 5, 120

\bibitem[{Al{\'i}-Lagoa {et~al.}(2013)Al{\'i}-Lagoa, {\mbox{de~Le\'on}},
  Licandro, Delbo', Campins, Pinilla-Alonso, \& Kelley}]{Ali-Lagoa2013}
Al{\'i}-Lagoa, V., {\mbox{de~Le\'on}}, J., Licandro, J., {et~al.} 2013, \aap,
  554, A71

\bibitem[{Ansdell {et~al.}(2014)Ansdell, Meech, Hainaut, Buie, Kaluna, Bauer,
  \& Dundon}]{ansdell2014}
Ansdell, M., Meech, K.~J., Hainaut, O., {et~al.} 2014, \apj, 793, 50

\bibitem[{Arai {et~al.}(2012)Arai, Kasuga, \& Ohtsuka}]{arai2012}
Arai, T., Kasuga, T., \& Ohtsuka, K. 2012, LPI Contributions, 1667, 6220

\bibitem[{Babadzhanov \& Obrubov(1992)}]{babadzhanov1992}
Babadzhanov, P.~B., \& Obrubov, {\mbox{Yu}}.~V. 1992, Celes. Mech. Dyn.
  Astron., 54, 111

\bibitem[{Belton \& A'Hearn(1999)}]{belton1999}
Belton, M. J.~S., \& A'Hearn, M.~F. 1999, Adv. Space Res., 24, 1167

\bibitem[{Binzel {et~al.}(2001)Binzel, Harris, Bus, \& Burbine}]{binzel2001}
Binzel, R.~P., Harris, A.~W., Bus, S.~J., \& Burbine, T.~H. 2001, \icarus, 151,
  139

\bibitem[{Binzel {et~al.}(2004)Binzel, Rivkin, Stuart, Harris, Bus, \&
  Burbine}]{binzel2004}
Binzel, R.~P., Rivkin, A.~S., Stuart, J.~S., {et~al.} 2004, \icarus, 170, 259

\bibitem[{Bus(1999)}]{Bus1999}
Bus, S.~J. 1999, PhD thesis, Massachusetts Institute of Technology, pp.~367

\bibitem[{Bus \& Binzel(2002)}]{bus2002a}
Bus, S.~J., \& Binzel, R.~P. 2002, \icarus, 158, 106

\bibitem[{Chamberlin {et~al.}(1996)Chamberlin, McFadden, Schulz, Schleicher, \&
  Bus}]{chamberlin1996}
Chamberlin, A.~B., McFadden, L.-A., Schulz, R., Schleicher, D.~G., \& Bus,
  S.~J. 1996, \icarus, 119, 173

\bibitem[{Chaumard {et~al.}(2012)Chaumard, Devouard, Delbo, Provost, \&
  Zanda}]{chaumard2012}
Chaumard, N., Devouard, B., Delbo, M., Provost, A., \& Zanda, B. 2012, \icarus,
  220, 65

\bibitem[{Clark {et~al.}(2010)Clark, Ziffer, Nesvorn\'y, Campins, Rivkin,
  Hiroi, Barucci, Fulchignoni, Binzel, Fornasier, DeMeo, Ockert-Bell, Licandro,
  \& Moth\'{e}-Diniz}]{clark2010}
Clark, B.~E., Ziffer, J., Nesvorn\'y, D., {et~al.} 2010, J. Geophys. Res., 115,
  E06005

\bibitem[{Cloutis {et~al.}(2012{\natexlab{a}})Cloutis, Hudon, Hiroi, \&
  Gaffey}]{cloutis2012c}
Cloutis, E.~A., Hudon, P., Hiroi, T., \& Gaffey, M.~J. 2012{\natexlab{a}},
  \icarus, 221, 911

\bibitem[{Cloutis {et~al.}(2013)Cloutis, Hudon, Hiroi, Gaffey, Mann, Alexander,
  Bell, \& Clark}]{cloutis2013}
Cloutis, E.~A., Hudon, P., Hiroi, T., {et~al.} 2013, LPI Contributions, 1719,
  1550

\bibitem[{Cloutis {et~al.}(2012{\natexlab{b}})Cloutis, Hudon, Hiroi, Gaffey,
  Mann, \& Bell}]{cloutis2012b}
---. 2012{\natexlab{b}}, \icarus, 221, 328

\bibitem[{Cloutis {et~al.}(2010)Cloutis, Hudon, Romanek, Bishop, Reddy, Gaffey,
  \& Hardersen}]{cloutis2010}
Cloutis, E.~A., Hudon, P., Romanek, C.~S., {et~al.} 2010, Meteor. Planet. Sci.,
  45, 1668

\bibitem[{Cochran \& Barker(1984)}]{cochran1984}
Cochran, A.~L., \& Barker, E.~S. 1984, \icarus, 59, 296

\bibitem[{Delbo {et~al.}(2014)Delbo, Libourel, Wilkerson, Murdoch, Michel,
  Ramesh, Ganino, Verati, \& Marchi}]{delbo2014}
Delbo, M., Libourel, G., Wilkerson, J., {et~al.} 2014, \nat, 508, 233

\bibitem[{{\mbox{de~Le\'on}} {et~al.}(2010){\mbox{de~Le\'on}}, Campins,
  Tsiganis, Morbidelli, \& Licandro}]{deleon2010}
{\mbox{de~Le\'on}}, J., Campins, H., Tsiganis, K., Morbidelli, A., \& Licandro,
  J. 2010, \aap, 513, A26

\bibitem[{Dermawan(2004)}]{dermawan2004}
Dermawan, B. 2004, PhD thesis, School of Science, University of Tokyo, pp.~118

\bibitem[{Dundon(2005)}]{dundon2005}
Dundon, L. 2005, Master's thesis, University of Hawaii, pp.~66

\bibitem[{Fujiwara \& Tsukamoto(1981)}]{fujiwara1981}
Fujiwara, A., \& Tsukamoto, A. 1981, \icarus, 48, 329

\bibitem[{Gaftonyuk {et~al.}(2010)Gaftonyuk, Krugly, \&
  Molotov}]{gaftonyuk2010}
Gaftonyuk, N.~M., Krugly, {\mbox{Yu}}.~N., \& Molotov, I.~E. 2010, in
  Protecting the Earth against Collisions with Asteroids and Comet Nuclei,
  Proceedings of the International Conference ``Asteroid-Comet Hazard-2009'',
  ed. A.~M. Finkelstein, W.~F. Huebner, \& V.~A. Shor (St. Petersburg: Nauka),
  49--51, held in St. Petersburg, Russia, September 21--25, 2009

\bibitem[{Galushina {et~al.}(2015)Galushina, Ryabova, \&
  Skripnichenko}]{galushina2015}
Galushina, T.~{\mbox{Yu}}., Ryabova, G.~O., \& Skripnichenko, P.~V. 2015,
  Planet. Space Sci., 118, 296

\bibitem[{Gladman {et~al.}(1997)Gladman, Migliorini, Morbidelli, Zappal\`a,
  Michel, Cellino, Froeschl\'e, Levison, Bailey, \& Duncan}]{gladman1997}
Gladman, B., Migliorini, F., Morbidelli, A., {et~al.} 1997, Science, 277, 197

\bibitem[{Granvik {et~al.}(2016)Granvik, Morbidelli, Jedicke, Bolin, Bottke,
  Beshore, Vokrouhlick\'y, Delb\`o, \& Michel}]{granvik2016}
Granvik, M., Morbidelli, A., Jedicke, R., {et~al.} 2016, \nat, 530, 303

\bibitem[{Green \& Kowal(1983)}]{green1983}
Green, S.~F., \& Kowal, C. 1983, \iaucirc, 3878, {\#}1

\bibitem[{Greenwood {et~al.}(2010)Greenwood, Franchi, Kearsley, \&
  Alard}]{greenwood2010}
Greenwood, R.~C., Franchi, I.~A., Kearsley, A.~T., \& Alard, O. 2010, Geochim.
  Cosmochim. Acta, 74, 1684

\bibitem[{Hamane {et~al.}(2002)Hamane, Kawakita, Kinugasa, Yamamuro, \&
  Takeyama}]{hamane2002}
Hamane, T., Kawakita, H., Kinugasa, K., Yamamuro, T., \& Takeyama, N. 2002,
  Publ. Astron. Soc. Japan, 54, L35

\bibitem[{Hanu\v{s} {et~al.}(2016{\natexlab{a}})Hanu\v{s}, Delbo',
  Vokrouhlick\'y, Pravec, Emery, Al\'i-Lagoa, Bolin, Devog\`ele, Dyvig,
  Gal\'ad, Jedicke, Korno\v{s}, Ku\v{s}nir\'ak, Licandro, Reddy,
  {\mbox{-P}}~Rivet, Vil\'agi, \& Warner}]{hanus2016a}
Hanu\v{s}, J., Delbo', M., Vokrouhlick\'y, D., {et~al.} 2016{\natexlab{a}},
  \aap, 592, A34

\bibitem[{Hanu\v{s} {et~al.}(2016{\natexlab{b}})Hanu\v{s}, Delbo',
  Vokrouhlick\'y, Pravec, Emery, Al\'i-Lagoa, Bolin, Devog\`ele, Dyvig,
  Gal\'ad, Jedicke, Korno\v{s}, Ku\v{s}nir\'ak, Licandro, Reddy, Warner,
  {\mbox{-P}}~Rivet, \& Vil\'agi}]{hanus2016b}
---. 2016{\natexlab{b}}, DPS meeting, 48, 516.08

\bibitem[{Harris(1998)}]{harris1998}
Harris, A.~W. 1998, \icarus, 131, 291

\bibitem[{Hicks {et~al.}(1998)Hicks, Fink, \& Grundy}]{hicks1998}
Hicks, M.~D., Fink, U., \& Grundy, W.~M. 1998, \icarus, 133, 69

\bibitem[{Hiroi {et~al.}(1993)Hiroi, Pieters, Zolensky, \&
  Lipschutz}]{hiroi1993}
Hiroi, T., Pieters, C.~M., Zolensky, M.~E., \& Lipschutz, M.~E. 1993, Science,
  261, 1016

\bibitem[{Hiroi {et~al.}(1996)Hiroi, Zolensky, Pieters, \&
  Lipschutz}]{hiroi1996}
Hiroi, T., Zolensky, M.~E., Pieters, C.~M., \& Lipschutz, M.~E. 1996, Meteor.
  Planet. Sci., 31, 321

\bibitem[{Hsieh \& Jewitt(2005)}]{hsieh2005}
Hsieh, H., \& Jewitt, D. 2005, \apj, 624, 1093

\bibitem[{Hui \& Li(2017)}]{hui2017}
Hui, M.-T., \& Li, J. 2017, \aj, 153, 23

\bibitem[{Ito \& Malhotra(2010)}]{ito2010}
Ito, T., \& Malhotra, R. 2010, \aap, 519, A63

\bibitem[{Jenniskens {et~al.}(2009)Jenniskens, Shaddad, Numan, Elsir, Kudoda,
  Zolensky, Le, Robinson, Friedrich, Rumble, Steele, Chesley, Fitzsimmons,
  Duddy, Hsieh, Ramsay, Brown, Edwards, Tagliaferri, Boslough, Spalding,
  Dantowitz, Kozubal, Pravec, Borovicka, Charvat, Vaubaillon, Kuiper, Albers,
  Bishop, Mancinelli, Sandford, Milam, Nuevo, \& Worden}]{jenniskens2009}
Jenniskens, P., Shaddad, M.~H., Numan, D., {et~al.} 2009, \nat, 485, 485

\bibitem[{Jewitt(2013)}]{jewitt2013a}
Jewitt, D. 2013, \aj, 145, 133

\bibitem[{Jewitt \& Hsieh(2006)}]{jewitt2006}
Jewitt, D., \& Hsieh, H. 2006, \aj, 132, 1624

\bibitem[{Jewitt {et~al.}(2015)Jewitt, Hsieh, \& Agarwal}]{jewitt2015}
Jewitt, D., Hsieh, H., \& Agarwal, J. 2015, in Asteroids IV, ed. P.~Michel,
  F.~E. Demeo, \& W.~F. Bottke (Tucson, Arizona: The University of Arizona
  Press), 221--241

\bibitem[{Jewitt \& Li(2010)}]{jewitt2010}
Jewitt, D., \& Li, J. 2010, \aj, 140, 1519

\bibitem[{Jewitt {et~al.}(2013)Jewitt, Li, \& Agarwal}]{jewitt2013b}
Jewitt, D., Li, J., \& Agarwal, J. 2013, \apjl, 771, L36

\bibitem[{Kaasalainen {et~al.}(2005)Kaasalainen, Kaasalainen, \&
  Piironen}]{kaasalainen2005}
Kaasalainen, S., Kaasalainen, M., \& Piironen, J. 2005, \aap, 440, 1177

\bibitem[{Kadono {et~al.}(2009)Kadono, Arakawa, Ito, \& Ohtsuki}]{kadono2009}
Kadono, T., Arakawa, M., Ito, T., \& Ohtsuki, K. 2009, \icarus, 200, 694

\bibitem[{Kasuga(2009)}]{kasuga2009}
Kasuga, T. 2009, Earth, Moon, and Planets, 105, 321

\bibitem[{Kasuga \& Jewitt(2008)}]{kasuga2008}
Kasuga, T., \& Jewitt, D. 2008, \aj, 136, 881

\bibitem[{Kasuga {et~al.}(2006)Kasuga, Watanabe, \& Sato}]{kasuga2006}
Kasuga, T., Watanabe, J.-I., \& Sato, M. 2006, \mnras, 373, 1107

\bibitem[{Kinoshita {et~al.}(2007)Kinoshita, Ohtsuka, Sekiguchi, Watanabe, Ito,
  Arakida, Kasuga, Miyasaka, Nakamura, \& Lin}]{kinoshita2007}
Kinoshita, D., Ohtsuka, K., Sekiguchi, T., {et~al.} 2007, \aap, 466, 1153

\bibitem[{Krugly {et~al.}(2002)Krugly, Belskaya, Shevchenko, Chiorny, Velichko,
  Mottola, Erikson, Hahn, Nathues, Neukum, Gaftonyuk, \& Dotto}]{krugly2002}
Krugly, {\mbox{Yu}}.~N., Belskaya, I.~N., Shevchenko, V.~G., {et~al.} 2002,
  \icarus, 158, 294

\bibitem[{Krugly {et~al.}(2007)Krugly, Gaftonyuk, Belskaya, Chiorny,
  Shevchenko, Velichko, Lupishko, Konovalenko, Falkovich, \&
  Molotov}]{krugly2007}
Krugly, {\mbox{Yu}}.~N., Gaftonyuk, N.~M., Belskaya, I.~N., {et~al.} 2007, in
  Proceedings of the IAU Symposium No.~236, Vol. 236, Near Earth Objects, our
  Celestial Neighbors: Opportunity and Risk, ed. A.~Milani, G.~B. Valsecchi, \&
  D.~Vokrouhlick\'y (Cambridge: Cambridge University Press), 385--390, held in
  14--18 August 2006, Praha, Czech Republic

\bibitem[{Lauretta {et~al.}(2015)Lauretta, Bartels, Barucci, Bierhaus, Binzel,
  Bottke, Campins, Chesley, Clark, Clark, Cloutis, Connolly, Crombie, Delbo,
  Dworkin, Emery, Glavin, Hamilton, Hergenrother, Johnson, Keller, Michel,
  Nolan, Sandford, Scheeres, Simon, Sutter, Vokrouhlicky, \&
  Walsh}]{lauretta2015}
Lauretta, D.~S., Bartels, A.~E., Barucci, M.~A., {et~al.} 2015, Meteor. Planet.
  Sci., 50, 834

\bibitem[{Lazzarin {et~al.}(1996)Lazzarin, Barucci, \&
  Doressoundiram}]{lazzarin1996}
Lazzarin, M., Barucci, M.~A., \& Doressoundiram, A. 1996, \icarus, 122, 122

\bibitem[{Lebofsky \& Spencer(1989)}]{lebofsky1989}
Lebofsky, L.~A., \& Spencer, J.~R. 1989, in Asteroids II, ed. R.~P. Binzel,
  T.~Gehrels, \& M.~S. Matthews (Tucson, Arizona: The University of Arizona
  Press), 128--147

\bibitem[{Li \& Jewitt(2013)}]{li2013}
Li, J., \& Jewitt, D. 2013, \aj, 145, 154

\bibitem[{Licandro {et~al.}(2007)Licandro, Campins, Moth\'e-Diniz,
  Pinilla-Alonso, \& {\mbox{de~Le\'on}}}]{licandro2007}
Licandro, J., Campins, H., Moth\'e-Diniz, T., Pinilla-Alonso, N., \&
  {\mbox{de~Le\'on}}, J. 2007, \aap, 461, 751

\bibitem[{Luu \& Jewitt(1990)}]{luu1990}
Luu, J.~X., \& Jewitt, D.~C. 1990, \aj, 99, 1985

\bibitem[{Madiedo {et~al.}(2013)Madiedo, Trigo-Rodr{\'i}guez, Castro-Tirado,
  Ortiz, \& Cabrera-Ca\~no}]{madiedo2013}
Madiedo, J.~M., Trigo-Rodr{\'i}guez, J.~M., Castro-Tirado, A.~J., Ortiz, J.~L.,
  \& Cabrera-Ca\~no, J. 2013, \mnras, 436, 2818

\bibitem[{Mann {et~al.}(2004)Mann, Kimura, Biesecker, Tsurutani, Gr{\"u}n,
  McKibben, Liou, MacQueen, Mukai, Guhathakurta, \& Lamy}]{mann2004}
Mann, I., Kimura, H., Biesecker, D.~A., {et~al.} 2004, \ssr, 110, 269

\bibitem[{Matsuoka {et~al.}(2015)Matsuoka, Nakamura, Kimura, Hiroi, Nakamura,
  Okumura, \& Sasaki}]{matsuoka2015}
Matsuoka, M., Nakamura, T., Kimura, Y., {et~al.} 2015, \icarus, 254, 135

\bibitem[{Meech {et~al.}(1996)Meech, Hainaut, Buie, {\mbox{A'Hearn}}, \&
  Lisse}]{meech1996}
Meech, K.~J., Hainaut, O.~R., Buie, M.~W., {\mbox{A'Hearn}}, M., \& Lisse, C.
  1996, in Abstracts of Asteroids, Comets, and Meteors 2016, held in
  Versailles, France, 8--12 July, 1996

\bibitem[{Moroz {et~al.}(2004)Moroz, Baratta, Strazzulla, Starukhina, Dotto,
  Barucci, Arnold, \& Distefano}]{moroz2004}
Moroz, L., Baratta, G., Strazzulla, G., {et~al.} 2004, \icarus, 170, 214

\bibitem[{Moth{\'e}-Diniz {et~al.}(2008)Moth{\'e}-Diniz, Carvano, Bus, Duffard,
  \& Burbine}]{mothe-diniz2008}
Moth{\'e}-Diniz, T., Carvano, J.~M., Bus, S.~J., Duffard, R., \& Burbine, T.~H.
  2008, \icarus, 195, 277

\bibitem[{Nakamura(2005)}]{nakamura2005}
Nakamura, T. 2005, J. Miner. Petro. Sci., 100, 260

\bibitem[{Nakamura(2006)}]{nakamura2006}
---. 2006, Earth Planet. Sci. Lett., 242, 26

\bibitem[{Nakato {et~al.}(2008)Nakato, Nakamura, Kitajima, \&
  Noguchi}]{nakato2008}
Nakato, A., Nakamura, T., Kitajima, F., \& Noguchi, T. 2008, Earth, Planet, and
  Space, 60, 855

\bibitem[{Nozaki {et~al.}(2006)Nozaki, Nakamura, , \& Noguchi}]{nozaki2006}
Nozaki, W., Nakamura, T., , \& Noguchi, T. 2006, Meteor. Planet. Sci., 41, 1095

\bibitem[{Ohtsuka {et~al.}(2008)Ohtsuka, Arakida, Ito, Yoshikawa, \&
  Asher}]{ohtsuka2008}
Ohtsuka, K., Arakida, H., Ito, T., Yoshikawa, M., \& Asher, D.~J. 2008, Meteor.
  Planet. Sci. Supplement, 43, 5055

\bibitem[{Ohtsuka {et~al.}(2009)Ohtsuka, Nakato, Nakamura, Kinoshita, Ito,
  Yoshikawa, \& Hasegawa}]{ohtsuka2009}
Ohtsuka, K., Nakato, A., Nakamura, T., {et~al.} 2009, Publ. Astron. Soc. Japan,
  61, 1375

\bibitem[{Ohtsuka {et~al.}(2006)Ohtsuka, Sekiguchi, Kinoshita, Watanabe, Ito,
  Arakida, \& Kasuga}]{ohtsuka2006}
Ohtsuka, K., Sekiguchi, T., Kinoshita, D., {et~al.} 2006, \aap, 450, L25

\bibitem[{Ohtsuka {et~al.}(1997)Ohtsuka, Shimoda, Yoshikawa, \&
  Watanabe}]{ohtsuka1997}
Ohtsuka, K., Shimoda, C., Yoshikawa, M., \& Watanabe, J.-I. 1997, Earth, Moon,
  and Planets, 77, 83

\bibitem[{Opeil {et~al.}(2010)Opeil, Consolmagno, \& Britt}]{opeil2010}
Opeil, C.~P., Consolmagno, G.~J., \& Britt, D.~T. 2010, \icarus, 208, 449

\bibitem[{Pravec {et~al.}(1998)Pravec, Wolf, \& \v{S}arounov\'a}]{pravec1998}
Pravec, P., Wolf, M., \& \v{S}arounov\'a, L. 1998, \icarus, 136, 124

\bibitem[{Sarli {et~al.}(2015)Sarli, Kawakatsu, \& Arai}]{sarli2015}
Sarli, B.~V., Kawakatsu, Y., \& Arai, T. 2015, J. Spacecraft and Rockets, 52,
  739

\bibitem[{Tholen(1985)}]{tholen1985}
Tholen, D.~J. 1985, \iaucirc, 4034, {\#}2

\bibitem[{Tomeoka {et~al.}(2003)Tomeoka, Kiriyama, Nakamura, Yamahana, \&
  Sekine}]{tomeoka2003}
Tomeoka, K., Kiriyama, K., Nakamura, K., Yamahana, Y., \& Sekine, T. 2003,
  \nat, 423, 60

\bibitem[{Urakawa {et~al.}(2014)Urakawa, Ohtsuka, Abe, Ito, \&
  Nakamura}]{urakawa2014}
Urakawa, S., Ohtsuka, K., Abe, S., Ito, T., \& Nakamura, T. 2014, \aj, 147, 121

\bibitem[{Urakawa {et~al.}(2002)Urakawa, Takahashi, Fujii, Ishiguro, Muka, \&
  Nakamura}]{urakawa2002}
Urakawa, S., Takahashi, S., Fujii, Y., {et~al.} 2002, in Dust in the Solar
  System and Other Planetary Systems, ed. S.~F. Green, I.~Williams,
  T.~McDonneli, \& N.~McBride (Oxford: Pergamon), 83--86, proceedings of the
  IAU Colloquium 181 held at the University of Kent, Canterbury, UK, 4--10
  April 2000 (COSPAR)

\bibitem[{Vilas \& Gaffey(1989)}]{vilas1989}
Vilas, F., \& Gaffey, M.~J. 1989, Science, 246, 790

\bibitem[{Warner(2015)}]{warner2015}
Warner, B.~D. 2015, Minor Planet Bulletin, 42, 115

\bibitem[{Whipple(1983)}]{whipple1983}
Whipple, F.~L. 1983, \iaucirc, 3881, {\#}1

\bibitem[{Zappal\`a(1980)}]{zappala1980}
Zappal\`a, V. 1980, The Moon and the Planets, 23, 345

\end{thebibliography}

\clearpage

\begin{table*}[htb]
\caption{Specification of the spectrograph Hiyoyu}
\label{hiyoyu}
\centering
\begin{tabular}{ll}
\hline
\hline
collimating lens & focal length $= 240$ mm  \\
camera lens & focal length $= 200$ mm \\
dispersion element & reflective grating \\
                   & 300 gr mm$^{-1}$ (or 1200 gr mm$^{-1}$) \\
slit size & $390'' \times 1.5''$ (projected on the sky) \\
detector & 1K-CCD Apogee AP-8 (Kodak KAF-1001E) \\ 
spectral resolution $R(\lambda/\Delta \lambda)$ & $\sim 333.133$ at $\lambda = 5000$ \AA \\ 
spectral wavelength range & 3800--7600 \AA \, (or 5000--9000 \AA) \\
wavelength calibration lamp & Fe-Ne-Ar hollow cathode lamp \\
slit viewer & SBIG ST-8 (for target acquisition and autoguiding) \\
\hline
\end{tabular}
\end{table*}

\clearpage

\begin{deluxetable}{cccccc}
 \tabletypesize{\footnotesize}
 \tablewidth{0pt}
 \tablecaption{Log of the spectroscopic observations of Phaethon}
 \tablehead{
 \colhead{Sp ID} & \colhead{Date} & \colhead{Midtime} & \colhead{Frame ID} & \colhead{Exp.} &
 \colhead{AM} \\
 \colhead{} & \colhead{UT 2007} & \colhead{UT} & \colhead{} & \colhead{sec.} & 
 \colhead{}
 }
\startdata
1130--01 & Nov 30 & 17:51:01 & 3200Phaethon-2091obj & 300 & 1.36 \\
1130--02 & Nov 30 & 17:58:12 & 3200Phaethon-2092obj & 300 & 1.33 \\
1130--03 & Nov 30 & 18:04:24 & 3200Phaethon-2093obj & 300 & 1.30 \\
1130--04 & Nov 30 & 18:10:27 & 3200Phaethon-2094obj & 300 & 1.28 \\
1130--05 & Nov 30 & 18:16:50 & 3200Phaethon-2095obj & 300 & 1.25 \\
1130--06 & Nov 30 & 18:22:58 & 3200Phaethon-2096obj & 300 & 1.23 \\
1130--07 & Nov 30 & 18:36:08 & 3200Phaethon-2099obj & 300 & 1.19 \\
1130--08 & Nov 30 & 18:42:44 & 3200Phaethon-2100obj & 300 & 1.17 \\
1130--09 & Nov 30 & 18:50:15 & 3200Phaethon-2101obj & 300 & 1.15 \\
1130--10 & Nov 30 & 18:56:45 & 3200Phaethon-2102obj & 300 & 1.13 \\
1130--11 & Nov 30 & 19:03:09 & 3200Phaethon-2103obj & 300 & 1.12 \\
1130--12 & Nov 30 & 19:10:02 & 3200Phaethon-2104obj & 300 & 1.10 \\
1130--13 & Nov 30 & 19:24:40 & 3200Phaethon-2107obj & 300 & 1.08 \\
1130--14 & Nov 30 & 19:30:51 & 3200Phaethon-2108obj & 300 & 1.07 \\
1130--15 & Nov 30 & 19:36:58 & 3200Phaethon-2109obj & 300 & 1.06 \\
1130--16 & Nov 30 & 19:43:14 & 3200Phaethon-2110obj & 300 & 1.05 \\
1130--17 & Nov 30 & 19:49:35 & 3200Phaethon-2111obj & 300 & 1.04 \\
1130--18 & Nov 30 & 19:55:43 & 3200Phaethon-2112obj & 300 & 1.03 \\
1130--19 & Nov 30 & 20:16:55 & 3200Phaethon-2115obj & 300 & 1.02 \\
1130--20 & Nov 30 & 20:23:08 & 3200Phaethon-2116obj & 300 & 1.01 \\
1130--21 & Nov 30 & 20:29:07 & 3200Phaethon-2117obj & 300 & 1.01 \\
1130--22 & Nov 30 & 20:35:07 & 3200Phaethon-2118obj & 300 & 1.01 \\
1130--23 & Nov 30 & 20:41:16 & 3200Phaethon-2119obj & 300 & 1.00 \\
1130--24 & Nov 30 & 20:47:28 & 3200Phaethon-2120obj & 300 & 1.00 \\
%
1201--01 & Dec 01 & 17:10:42 & 3200Phaethon-2067obj & 300 & 1.68 \\
1201--02 & Dec 01 & 17:17:24 & 3200Phaethon-2068obj & 300 & 1.63 \\
1201--03 & Dec 01 & 17:24:04 & 3200Phaethon-2069obj & 300 & 1.57 \\
1201--04 & Dec 01 & 17:30:55 & 3200Phaethon-2070obj & 300 & 1.53 \\
1201--05 & Dec 01 & 17:37:26 & 3200Phaethon-2071obj & 300 & 1.49 \\
1201--06 & Dec 01 & 17:43:48 & 3200Phaethon-2072obj & 300 & 1.44 \\
1201--07 & Dec 01 & 18:03:41 & 3200Phaethon-2076obj & 300 & 1.34 \\ 
1201--08 & Dec 01 & 18:09:59 & 3200Phaethon-2077obj & 300 & 1.31 \\ 
1201--09 & Dec 01 & 18:16:13 & 3200Phaethon-2078obj & 300 & 1.29 \\ %
1201--10 & Dec 01 & 18:22:53 & 3200Phaethon-2079obj & 300 & 1.26 \\
1201--11 & Dec 01 & 18:34:42 & 3200Phaethon-2082obj & 300 & 1.22 \\
1201--12 & Dec 01 & 18:41:25 & 3200Phaethon-2083obj & 300 & 1.20 \\
1201--13 & Dec 01 & 18:48:23 & 3200Phaethon-2084obj & 300 & 1.18 \\
1201--14 & Dec 01 & 18:54:46 & 3200Phaethon-2085obj & 300 & 1.16 \\
1201--15 & Dec 01 & 19:01:25 & 3200Phaethon-2086obj & 300 & 1.14 \\
1201--16 & Dec 01 & 19:07:50 & 3200Phaethon-2087obj & 300 & 1.13 \\
1201--17 & Dec 01 & 19:35:50 & 3200Phaethon-2096obj & 300 & 1.07 \\
1201--18 & Dec 01 & 19:42:22 & 3200Phaethon-2097obj & 300 & 1.06 \\
1201--19 & Dec 01 & 19:48:58 & 3200Phaethon-2098obj & 300 & 1.05 \\
1201--20 & Dec 01 & 19:55:35 & 3200Phaethon-2099obj & 300 & 1.04 \\
1201--21 & Dec 01 & 20:01:47 & 3200Phaethon-2100obj & 300 & 1.04 \\
1201--22 & Dec 01 & 20:08:03 & 3200Phaethon-2101obj & 300 & 1.03 \\
1201--23 & Dec 01 & 20:38:09 & 3200Phaethon-2110obj & 300 & 1.01 \\
1201--24 & Dec 01 & 20:44:19 & 3200Phaethon-2111obj & 300 & 1.00 \\
1201--25 & Dec 01 & 20:51:15 & 3200Phaethon-2112obj & 300 & 1.00 \\
1201--26 & Dec 01 & 20:58:41 & 3200Phaethon-2113obj & 300 & 1.00 \\
1201--27 & Dec 01 & 21:04:41 & 3200Phaethon-2114obj & 300 & 1.00 \\
1202--01 & Dec 02 & 18:11:07 & 3200Phaethon-2047obj & 300 & 1.36 \\
1202--02 & Dec 02 & 18:23:17 & 3200Phaethon-2049obj & 300 & 1.30 \\ 
1202--03 & Dec 02 & 18:29:21 & 3200Phaethon-2050obj & 300 & 1.28 \\ 
1202--04 & Dec 02 & 18:35:28 & 3200Phaethon-2051obj & 300 & 1.26 \\
1202--05 & Dec 02 & 18:41:53 & 3200Phaethon-2052obj & 300 & 1.23 \\
1202--06 & Dec 02 & 19:13:40 & 3200Phaethon-2065obj & 300 & 1.14 \\
1202--07 & Dec 02 & 19:19:29 & 3200Phaethon-2066obj & 300 & 1.12 \\
1202--08 & Dec 02 & 19:25:38 & 3200Phaethon-2067obj & 300 & 1.11 \\
1202--09 & Dec 02 & 19:32:25 & 3200Phaethon-2068obj & 300 & 1.10 \\
1202--10 & Dec 02 & 19:38:54 & 3200Phaethon-2069obj & 300 & 1.08 \\
1202--11 & Dec 02 & 19:45:04 & 3200Phaethon-2070obj & 300 & 1.07 \\
1202--12 & Dec 02 & 20:13:00 & 3200Phaethon-2081obj & 300 & 1.04 \\
1202--13 & Dec 02 & 20:18:58 & 3200Phaethon-2082obj & 300 & 1.03 \\
1202--14 & Dec 02 & 20:25:17 & 3200Phaethon-2083obj & 300 & 1.02 \\
1202--15 & Dec 02 & 20:31:46 & 3200Phaethon-2084obj & 300 & 1.02 \\
1202--16 & Dec 02 & 20:37:53 & 3200Phaethon-2085obj & 300 & 1.01 \\
1202--17 & Dec 02 & 20:43:43 & 3200Phaethon-2086obj & 300 & 1.01 \\
\enddata
\begin{list}{}{}
\item[Note to the column heads.] $\,$ \\ 
Sp ID: spectrum ID, \\
Date and Midtime: observed date in 2007 and midtime of exposure (in UT), \\ 
Frame ID: Lulin spectral file ID in (hyyyymmdd\_)OBJ-NO.fits, \\ 
Exp.: exposure (in second), \\ 
AM: airmass. \\ 
\end{list}
\label{obslog}
\end{deluxetable}

\clearpage

\begin{deluxetable}{ccccccccr}
 \tabletypesize{\footnotesize}
 \tablewidth{0pt}
 \tablecaption{Results of the VIS-spectrophotometries of Phaethon along with their geometries}
 \tablehead{
 \colhead{Sp ID} & \colhead{JDT} & \colhead{$\Delta$} & \colhead{$r$} & \colhead{$\alpha$} & 
 \colhead{$\phi_{\rm K}$} & \colhead{$\phi_{\rm A}$} & \colhead{$\phi_{\rm H}$} & \colhead{$S'$} \\
 \colhead{} & \colhead{2454430+} & \colhead{AU} &  \colhead{AU} & \colhead{deg} & 
 \colhead{deg} & \colhead{deg} & \colhead{deg} & \colhead{$\%/0.1$ $\micron$} 
 }
\startdata
1130--01 & 5.24452 & 0.2357 & 1.0862 & 59.11 & $-43.22$ & $-28.88$ & $-61.86$ & $-6.70 \pm 0.53$ \\
1130--02 & 5.24951 & 0.2356 & 1.0862 & 59.13 & $-43.21$ & $-28.87$ & $-61.86$ & $-11.45 \pm 1.25$ \\
1130--03 & 5.25381 & 0.2356 & 1.0861 & 59.14 & $-43.20$ & $-28.86$ & $-61.86$ & $-7.86 \pm 0.86$ \\
1130--04 & 5.25801 & 0.2355 & 1.0860 & 59.15 & $-43.19$ & $-28.85$ & $-61.86$ & $-9.20 \pm 0.50$ \\
1130--05 & 5.26245 & 0.2354 & 1.0860 & 59.16 & $-43.18$ & $-28.84$ & $-61.86$ & $-7.49 \pm 0.70$ \\
1130--06 & 5.26671 & 0.2353 & 1.0859 & 59.17 & $-43.17$ & $-28.83$ & $-61.86$ & $-8.31 \pm 0.42$ \\
1130--07 & 5.27585 & 0.2352 & 1.0857 & 59.19 & $-43.14$ & $-28.81$ & $-61.86$ & $-6.82 \pm 1.56$ \\
1130--08 & 5.28043 & 0.2351 & 1.0857 & 59.20 & $-43.13$ & $-28.80$ & $-61.86$ & $-6.90 \pm 0.40$ \\
1130--09 & 5.28565 & 0.2350 & 1.0856 & 59.21 & $-43.12$ & $-28.79$ & $-61.86$ & $-7.91 \pm 0.57$ \\
1130--10 & 5.29017 & 0.2349 & 1.0855 & 59.22 & $-43.11$ & $-28.78$ & $-61.86$ & $-6.24 \pm 0.83$ \\
1130--11 & 5.29461 & 0.2348 & 1.0855 & 59.23 & $-43.10$ & $-28.77$ & $-61.86$ & $ 0.30 \pm 1.53$ \\
1130--12 & 5.29939 & 0.2347 & 1.0854 & 59.24 & $-43.09$ & $-28.75$ & $-61.86$ & $-3.23 \pm 0.91$ \\
1130--13 & 5.30955 & 0.2345 & 1.0852 & 59.27 & $-43.06$ & $-28.73$ & $-61.86$ & $-3.53 \pm 0.34$ \\
1130--14 & 5.31385 & 0.2345 & 1.0851 & 59.28 & $-43.05$ & $-28.72$ & $-61.86$ & $-7.17 \pm 0.66$ \\
1130--15 & 5.31810 & 0.2344 & 1.0851 & 59.29 & $-43.04$ & $-28.71$ & $-61.86$ & $-3.07 \pm 0.35$ \\
1130--16 & 5.32245 & 0.2343 & 1.0850 & 59.30 & $-43.03$ & $-28.70$ & $-61.86$ & $-2.31 \pm 0.73$ \\
1130--17 & 5.32686 & 0.2342 & 1.0849 & 59.31 & $-43.02$ & $-28.69$ & $-61.86$ & $-2.57 \pm 0.60$ \\
1130--18 & 5.33112 & 0.2341 & 1.0849 & 59.32 & $-43.01$ & $-28.68$ & $-61.86$ & $-4.24 \pm 0.62$ \\
1130--19 & 5.34584 & 0.2339 & 1.0846 & 59.35 & $-42.97$ & $-28.64$ & $-61.86$ & $-4.36 \pm 0.32$ \\
1130--20 & 5.35016 & 0.2338 & 1.0846 & 59.36 & $-42.96$ & $-28.63$ & $-61.86$ & $-0.00 \pm 0.30$ \\
1130--21 & 5.35431 & 0.2337 & 1.0845 & 59.37 & $-42.95$ & $-28.62$ & $-61.86$ & $-1.08 \pm 0.26$ \\
1130--22 & 5.35848 & 0.2336 & 1.0844 & 59.38 & $-42.94$ & $-28.61$ & $-61.86$ & $-2.64 \pm 0.48$ \\
1130--23 & 5.36275 & 0.2336 & 1.0844 & 59.39 & $-42.93$ & $-28.60$ & $-61.86$ & $-3.83 \pm 0.23$ \\
1130--24 & 5.36705 & 0.2335 & 1.0843 & 59.40 & $-42.92$ & $-28.59$ & $-61.86$ & $-2.99 \pm 0.22$ \\
1201--01 & 6.21652 & 0.2182 & 1.0709 & 61.62 & $-40.67$ & $-26.43$ & $-61.69$ & $-6.38 \pm 0.45$ \\
1201--02 & 6.22117 & 0.2181 & 1.0708 & 61.63 & $-40.66$ & $-26.42$ & $-61.69$ & $-10.26 \pm 0.45$ \\
1201--03 & 6.22580 & 0.2180 & 1.0708 & 61.64 & $-40.64$ & $-26.40$ & $-61.69$ & $-4.11 \pm 0.37$ \\
1201--04 & 6.23056 & 0.2179 & 1.0707 & 61.66 & $-40.63$ & $-26.39$ & $-61.69$ & $-3.90 \pm 0.34$ \\
1201--05 & 6.23509 & 0.2179 & 1.0706 & 61.67 & $-40.62$ & $-26.38$ & $-61.69$ & $-4.45 \pm 0.48$ \\
1201--06 & 6.23951 & 0.2178 & 1.0705 & 61.68 & $-40.61$ & $-26.37$ & $-61.69$ & $-5.25 \pm 0.34$ \\
1201--07 & 6.25332 & 0.2175 & 1.0703 & 61.72 & $-40.57$ & $-26.33$ & $-61.68$ & $-2.96 \pm 0.34$ \\
1201--08 & 6.25769 & 0.2175 & 1.0703 & 61.73 & $-40.55$ & $-26.32$ & $-61.68$ & $-1.33 \pm 0.41$ \\
1201--09 & 6.26202 & 0.2174 & 1.0702 & 61.74 & $-40.54$ & $-26.30$ & $-61.68$ & $ 0.63 \pm 0.43$ \\
1201--10 & 6.26665 & 0.2173 & 1.0701 & 61.76 & $-40.53$ & $-26.29$ & $-61.68$ & $-1.10 \pm 0.28$ \\
1201--11 & 6.27485 & 0.2172 & 1.0700 & 61.78 & $-40.50$ & $-26.27$ & $-61.67$ & $-0.26 \pm 0.32$ \\
1201--12 & 6.27952 & 0.2171 & 1.0699 & 61.79 & $-40.49$ & $-26.26$ & $-61.67$ & $-3.05 \pm 0.28$ \\
1201--13 & 6.28436 & 0.2170 & 1.0698 & 61.80 & $-40.48$ & $-26.24$ & $-61.67$ & $-2.40 \pm 0.33$ \\
1201--14 & 6.28879 & 0.2169 & 1.0698 & 61.82 & $-40.46$ & $-26.23$ & $-61.67$ & $-4.28 \pm 0.33$ \\
1201--15 & 6.29341 & 0.2168 & 1.0697 & 61.83 & $-40.45$ & $-26.22$ & $-61.67$ & $-3.69 \pm 0.31$ \\
1201--16 & 6.29786 & 0.2167 & 1.0696 & 61.84 & $-40.44$ & $-26.21$ & $-61.67$ & $-1.91 \pm 0.28$ \\
1201--17 & 6.31731 & 0.2164 & 1.0693 & 61.90 & $-40.38$ & $-26.15$ & $-61.66$ & $ 1.22 \pm 0.40$ \\
1201--18 & 6.32185 & 0.2163 & 1.0692 & 61.91 & $-40.37$ & $-26.14$ & $-61.66$ & $ 1.59 \pm 0.40$ \\
1201--19 & 6.32643 & 0.2162 & 1.0692 & 61.92 & $-40.36$ & $-26.13$ & $-61.66$ & $-0.02 \pm 1.96$ \\
1201--20 & 6.33102 & 0.2162 & 1.0691 & 61.94 & $-40.34$ & $-26.11$ & $-61.65$ & $ 2.67 \pm 0.35$ \\
1201--21 & 6.33533 & 0.2161 & 1.0690 & 61.95 & $-40.33$ & $-26.10$ & $-61.65$ & $ 1.58 \pm 0.39$ \\
1201--22 & 6.33968 & 0.2160 & 1.0690 & 61.96 & $-40.32$ & $-26.09$ & $-61.65$ & $ 1.82 \pm 0.36$ \\
1201--23 & 6.36058 & 0.2156 & 1.0686 & 62.02 & $-40.26$ & $-26.03$ & $-61.64$ & $-1.40 \pm 0.31$ \\
1201--24 & 6.36487 & 0.2156 & 1.0686 & 62.03 & $-40.24$ & $-26.02$ & $-61.64$ & $-1.60 \pm 1.01$ \\
1201--25 & 6.36968 & 0.2155 & 1.0685 & 62.04 & $-40.23$ & $-26.01$ & $-61.64$ & $ 0.36 \pm 0.32$ \\
1201--26 & 6.37484 & 0.2154 & 1.0684 & 62.06 & $-40.22$ & $-25.99$ & $-61.64$ & $ 0.44 \pm 0.25$ \\
1201--27 & 6.37901 & 0.2153 & 1.0683 & 62.07 & $-40.20$ & $-25.98$ & $-61.64$ & $ 0.62 \pm 0.32$ \\
1202--01 & 7.25848 & 0.2001 & 1.0543 & 64.81 & $-37.43$ & $-23.32$ & $-61.11$ & $ 5.23 \pm 0.54$ \\
1202--02 & 7.26693 & 0.1999 & 1.0541 & 64.84 & $-37.40$ & $-23.29$ & $-61.11$ & $ 4.37 \pm 1.45$ \\
1202--03 & 7.27114 & 0.1998 & 1.0541 & 64.85 & $-37.39$ & $-23.28$ & $-61.10$ & $ 3.95 \pm 0.43$ \\
1202--04 & 7.27539 & 0.1998 & 1.0540 & 64.87 & $-37.38$ & $-23.26$ & $-61.10$ & $ 5.86 \pm 0.29$ \\
1202--05 & 7.27984 & 0.1997 & 1.0539 & 64.88 & $-37.36$ & $-23.25$ & $-61.09$ & $ 6.63 \pm 0.26$ \\
1202--06 & 7.30191 & 0.1993 & 1.0536 & 64.96 & $-37.29$ & $-23.18$ & $-61.08$ & $-7.23 \pm 0.44$ \\
1202--07 & 7.30595 & 0.1993 & 1.0535 & 64.97 & $-37.27$ & $-23.16$ & $-61.07$ & $-6.44 \pm 0.41$ \\
1202--08 & 7.31023 & 0.1992 & 1.0534 & 64.98 & $-37.26$ & $-23.15$ & $-61.07$ & $-8.05 \pm 0.34$ \\
1202--09 & 7.31494 & 0.1991 & 1.0534 & 65.00 & $-37.24$ & $-23.13$ & $-61.07$ & $-8.35 \pm 0.43$ \\
1202--10 & 7.31944 & 0.1990 & 1.0533 & 65.01 & $-37.23$ & $-23.12$ & $-61.06$ & $-6.28 \pm 0.51$ \\
1202--11 & 7.32372 & 0.1990 & 1.0532 & 65.03 & $-37.21$ & $-23.11$ & $-61.06$ & $-6.22 \pm 0.27$ \\
1202--12 & 7.34312 & 0.1986 & 1.0529 & 65.09 & $-37.14$ & $-23.04$ & $-61.04$ & $-5.67 \pm 0.26$ \\
1202--13 & 7.34726 & 0.1986 & 1.0528 & 65.11 & $-37.13$ & $-23.03$ & $-61.04$ & $-7.35 \pm 0.22$ \\
1202--14 & 7.35165 & 0.1985 & 1.0528 & 65.12 & $-37.11$ & $-23.01$ & $-61.03$ & $-6.40 \pm 0.27$ \\
1202--15 & 7.35615 & 0.1984 & 1.0527 & 65.14 & $-37.10$ & $-23.00$ & $-61.03$ & $-6.89 \pm 0.26$ \\
1202--16 & 7.36040 & 0.1983 & 1.0526 & 65.15 & $-37.08$ & $-22.98$ & $-61.03$ & $-4.23 \pm 1.83$ \\
1202--17 & 7.36445 & 0.1983 & 1.0526 & 65.17 & $-37.07$ & $-22.97$ & $-61.02$ & $-5.78 \pm 0.39$ \\
\enddata
\begin{list}{}{}
\item[Note to the column heads.] $\,$ \\
Sp ID: spectrum ID, \\
JDT: Julian Terrestrial Date, \\
$\Delta$ and $r$: geocentric and heliocentric distances (in AU), \\ 
$\alpha$: phase angle (in degree), \\
$\phi_{\rm K}$, $\phi_{\rm A}$, and $\phi_{\rm H}$: sub-Earth latitudes (in degree), based on the pole solutions, SOL K by \citet{krugly2002}, SOL A by \citet{ansdell2014}, and SOL H by \citet{hanus2016a}, respectively, \\
$S'$: spectral gradient (in $\%/0.1$ $\micron$). \\
\end{list}
\label{obsre}
\end{deluxetable}

\clearpage

\begin{deluxetable}{lccccl}
 \tabletypesize{\footnotesize}
 \tablewidth{0pt}
 \tablecaption{Pole axis parameters of Phaethon's rotation for the geometric analyses}
 \tablehead{
 \colhead{Solution} & \colhead{$\lambda_{\rm P}$} & \colhead{$\beta_{\rm P}$} & \colhead{$P$} & \colhead{\# Nights} &
 \colhead{Arc} \\
 \colhead{ID} & \colhead{deg} & \colhead{deg} & \colhead{hr} & \colhead{} & \colhead{}
 }
\startdata
SOL K & $97 \pm 15$ & $-11\pm 15$ & $3.5904399 \pm 0.000002$&  6 & 1996 Sep--1997 Nov \\
SOL A & $85 \pm 13$ & $-20 \pm 10$ &  $3.6032 \pm 0.0008$ & 15 & 1994 Dec 27--2013 Dec 11\\
SOL H & $319\pm 5$  & $-39 \pm 5$ & $3.603958 \pm 0.000002$ & 49 & 1994 Nov 02--2015 Oct 08 \\
\enddata
\begin{list}{}{}
\item[References to solution ID.] $\,$ \\

SOL K \citep{krugly2002}, SOL A \citep{ansdell2014}, and SOL H \citep{hanus2016a}. \\ 
\item[Note to the column heads.] $\,$ \\
$\lambda_{\rm P}$ and $\beta_{\rm P}$: ecliptic longitude and latitude (in degree, equinox J2000) of pole axis orientation, \\  
$P$: sidereal rotation period (in hour), \\ 
\# Nights and Arc: number of observed night and its photometric arc, taken in the pole axis determination. \\ 
\end{list}
\label{psol}
\end{deluxetable}

\clearpage

\begin{deluxetable}{llllrrrrrcl}
 \tabletypesize{\footnotesize}
 \tablewidth{0pt}
 \tablecaption{Other VIS-spectroscopic data of Phaethon}
 \tablehead{
 \colhead{Sp ID} & \colhead{Date} & \colhead{$\Delta$} & \colhead{$r$} &
 \colhead{$\alpha$} & \colhead{$\phi_{\rm K}$} & \colhead{$\phi_{\rm A}$} & \colhead{$\phi_{\rm H}$} & \colhead{$S'$} & \colhead{EWR} & \colhead{Tel.} \\
 \colhead{} & \colhead{UT} & \colhead{AU} &  \colhead{AU} & 
 \colhead{deg} & \colhead{deg} & \colhead{deg} & \colhead{deg} & \colhead{} & \colhead{$\micron$} & \colhead{}
 }
\startdata
CB1983   & 1983 Dec 02     & 1.16 & 1.67 & 35.0 &   16.6 &  10.0   &   16.3  &   23.0  & 0.36--0.64 & HJS 2.7m     \\
LJ1988   & 1988 Oct 06     & 1.75 & 2.38 & 21.8 & $-53.8$& $-51.2$ & $-30.2$ & $-11.0$ & 0.36--0.72 & KPNO 2.1m    \\
Ch1993   & 1993 Oct 21     & 0.43 & 0.99 & 78.0 &   44.4 &   52.8  & $-12.8$ & $- 4.0$ & 0.36--0.62 & Perkins 1.8m \\
La1994   & 1994 Oct 10     & 1.49 & 1.99 & 29.0 & $-65.9$& $-57.0$ & $-40.3$ & $- 3.9$ & 0.38--1.00 & CFHT 3.6m    \\
8        & 1994 Nov 15     & 0.81 & 1.71 & 20.0 & $-61.0$& $-54.3$ & $-38.2$ & $- 1.7$ & 0.44--0.93 & MDM 2.4m     \\
Hi1995   & 1995 Nov 26     & 1.49 & 2.40 & 12.0 & $-27.6$& $-31.5$ & $- 8.3$ & $- 1.8$ & 0.55--0.95 & Kuiper 1.54m \\
sp19     & 2002 Oct 27     & 1.37 & 2.28 & 13.0 & $-31.8$& $-33.2$ & $-17.4$ & $- 2.1$ & 0.44--0.93 & MIT-SMASS    \\
Li2003   & 2003 Nov 14.8   & 0.72 & 1.18 & 56.9 &   47.3 &   42.8  &   16.4  & $- 4.0$ & 0.52--0.95 & WHT 4.2m     \\
sp34     & 2004 Dec 10     & 0.64 & 1.58 & 17.0 & $-40.3$& $-43.4$ & $-14.6$ & $- 2.2$ & 0.44--0.93 & MIT-SMASS    \\
Li2004a  & 2004 Dec 21.0   & 0.61 & 1.46 & 30.9 & $-29.5$& $-35.5$ & $- 2.5$ & $- 4.0$ & 0.35--0.60 & NOT 2.5m     \\
Li2004b  & 2004 Dec 21.0   & 0.61 & 1.46 & 30.9 & $-29.5$& $-35.5$ & $- 2.5$ & $- 4.0$ & 0.60--0.90 & NOT 2.5m     \\
sp204n1  & 2014 Nov 28     & 0.85 & 1.81 &  9.3 & $-47.5$& $-47.8$ & $-23.0$ & $- 2.1$ & 0.44--0.93 & MIT-SMASS    \\
sp204n2  & 2014 Nov 29     & 0.84 & 1.80 &  9.4 & $-46.8$& $-47.4$ & $-22.2$ & $- 2.1$ & 0.44--0.93 & MIT-SMASS    \\
dm19n2   & 2014 Dec 02     & 0.82 & 1.78 & 10.4 & $-44.7$& $-46.1$ & $-19.9$ & $- 2.1$ & 0.44--0.93 & MIT-SMASS    \\ 
4        &                 &      &      &      &        &       &       & $- 6.8$ & 0.36--0.93 & Palomar 5m   \\

\enddata
\begin{list}{}{}
\item[References to Sp ID (spectrum ID).] $\,$ \\
CB1983 \citep{cochran1984}, LJ1988 \citep{luu1990}, Ch1996 \citep{chamberlin1996}, La1994 \citep{lazzarin1996},  
8 \citep{binzel2004}, Hi1995 \citep{hicks1998},   
sp19, 34, 204n1, 204n2, and dm19n2 (MIT--UH--IRTF Joint Campaign for NEO Reconnaissance, 
see http://smass.mit.edu/smass.html),   
Li2003, 2004a, and 2004b \citep{licandro2007},  
and 4 \citep[][whose time data were not given in]{binzel2001}. \\ 
\item[Note to the column heads.] $\,$ \\
Date: date and time of the observation (in UT), \\ 
$\Delta$ and $r$: geocentric and heliocentric distances (in AU), \\ 
$\alpha$: phase angle (in degree), \\
$\phi_{\rm K}$, $\phi_{\rm A}$, and $\phi_{\rm H}$: sub-Earth latitudes (in degree), based on SOLs K, A, and H, respectively, \\
$S'$: spectral gradient (in $\%/0.1$ $\micron$), \\
EWR: effective wavelength range of the spectrum (in $\micron$), \\
Tel.: telescope used in the observation. \\ 
\end{list}
\label{spdata}
\end{deluxetable}

\clearpage

\begin{deluxetable}{lllrcrl}
 \tabletypesize{\footnotesize}
 \tablewidth{0pt}
 \tablecaption{VIS-spectroscopic data of metamorphosed carbonaceous chondrites from the RELAB database}
 \tablehead{
 \colhead{Name} & \colhead{Type} & \colhead{Sample ID} &
 \colhead{Heated at} & \colhead{Gr. size} & \colhead{$S'$} & \colhead{Remark} \\
 \colhead{} & \colhead{} & \colhead{} &
 \colhead{$^\circ$C} & \colhead{$\micron$} & \colhead{$\%/0.1$ $\micron$} & \colhead{}
 }
\startdata
Y-82162   & CI-an    & MB-CMP-019-1 &      & $<125$ & $ 1.67 \pm 0.01$ &                   \\
          &          & MB-CMP-019-C &      &   chip & $-0.32 \pm 0.03$ & interior chip     \\
Ivuna     & CI1      & MP-TXH-018-F &  100 & $<125$ & $ 9.74 \pm 0.05$ & heated for 1 week \\
          &          & MP-TXH-018-G &  200 & $<125$ & $ 5.18 \pm 0.02$ & heated for 1 week \\
          &          & MP-TXH-018-A &  300 & $<125$ & $ 4.08 \pm 0.01$ & heated for 1 week \\
          &          & MP-TXH-018-B &  400 & $<125$ & $ 6.41 \pm 0.02$ & heated for 1 week \\
          &          & MP-TXH-018-C &  500 & $<125$ & $ 5.58 \pm 0.01$ & heated for 1 week \\
          &          & MP-TXH-018-D &  600 & $<125$ & $ 4.05 \pm 0.01$ & heated for 1 week \\
          &          & MP-TXH-018-E &  700 & $<125$ & $-4.24 \pm 0.02$ & heated for 1 week \\
Murchison & CM2      & MB-TXH-064-2 &      & $<125$ & $ 8.50 \pm 0.18$ &                   \\
          &          & MB-TXH-064-3 &      & $<63$  & $ 9.33 \pm 0.18$ &                   \\
          &          & MB-TXH-064-4 &      & 63--125& $ 6.21 \pm 0.14$ &                   \\
          &          & MB-TXH-064-A &  400 & $<63$  & $13.44 \pm 0.03$ & heated for 1 week \\
          &          & MB-TXH-064-B &  500 & $<63$  & $11.42 \pm 0.02$ & heated for 1 week \\
          &          & MB-TXH-064-C &  600 & $<63$  & $ 5.76 \pm 0.01$ & heated for 1 week \\
          &          & MB-TXH-064-D &  700 & $<63$  & $ 3.24 \pm 0.02$ & heated for 1 week \\
          &          & MB-TXH-064-E &  800 & $<63$  & $ 3.29 \pm 0.02$ & heated for 1 week \\
          &          & MB-TXH-064-F &  900 & $<63$  & $ 0.26 \pm 0.01$ & heated for 1 week \\
          &          & MB-TXH-064-G & 1000 & $<63$  & $ 1.44 \pm 0.02$ & heated for 1 week \\
          &          & MB-TXH-064-H &  400 & 63--125& $12.97 \pm 0.03$ & heated for 1 week \\
          &          & MB-TXH-064-I &  500 & 63--125& $ 9.81 \pm 0.03$ & heated for 1 week \\
          &          & MB-TXH-064-J &  600 & 63--125& $ 5.19 \pm 0.01$ & heated for 1 week \\
          &          & MB-TXH-064-K &  700 & 63--125& $ 1.84 \pm 0.01$ & heated for 1 week \\
          &          & MB-TXH-064-L &  800 & 63--125& $ 3.41 \pm 0.02$ & heated for 1 week \\
Murchison & CM2      & MB-TXH-064-M &  900 & 63--125& $-0.98 \pm 0.01$ & heated for 1 week \\
          &          & MB-TXH-064-N & 1000 & 63--125& $ 1.45 \pm 0.02$ & heated for 1 week \\
B-7904    & CM-an    & MB-CMP-018-1 &      &   chip & $ 6.82 \pm 0.04$ &                   \\
Y-86720   & CM-an    & MB-CMP-020-1 &      & $<125$ & $ 2.49 \pm 0.06$ &                   \\
          &          & MB-CMP-020-A &      &   chip & $ 3.21 \pm 0.09$ & smaller chip      \\
          &          & MB-CMP-020-B &      &   chip & $ 2.27 \pm 0.07$ & larger chip       \\
MET01070  & CM1	     & PH-D2M-043   &      & $<75$  & $ 0.43 \pm 0.08$ &                   \\
PCA02012  & CM2      & PH-D2M-044   &      & $<75$  & $28.85 \pm 0.59$ &                   \\
Y-693     & CK4      & MB-TXH-077   &      & $<125$ & $ 6.73 \pm 0.14$ &                   \\
ALH85002  & CK4      & MB-TXH-081-1 &      &   chip & $ 3.52 \pm 0.11$ &                   \\
          &          & MB-TXH-081-2 &      & $<125$ & $ 2.88 \pm 0.12$ &                   \\
          &          & MB-TXH-081-A &      & $<25$  & $ 5.18 \pm 0.08$ &                   \\
          &          & MB-TXH-081-B &      & 25--45 & $ 2.66 \pm 0.06$ &                   \\
          &          & MB-TXH-081-C &      & 45--75 & $ 0.34 \pm 0.08$ &                   \\
          &          & MB-TXH-081-D &      & 75--125& $ 1.31 \pm 0.06$ &                   \\
          &          & MH-FPF-058-A &      & $<180$ & $ 4.72 \pm 0.14$ &                   \\
          &          & MH-FPF-058-B &      & $<180$ & $ 7.29 \pm 0.19$ & fusion-crusted    \\
EET92002  & CK4      & MC-RPB-003   &      & $<500$ & $ 0.76 \pm 0.11$ &                   \\
MET01149  & CK3      & PH-D2M-045   &      & $<75$  & $ 7.46 \pm 0.26$ &                   \\
PCA91470  & CK4      & PH-D2M-046   &      & $<75$  & $ 3.94 \pm 0.22$ &                   \\
EET83311  & CK5      & PH-D2M-047   &      & $<75$  & $ 5.83 \pm 0.33$ &                   \\
DAV92300  & CK4      & PH-D2M-053   &      & $<75$  & $ 2.86 \pm 0.08$ &                   \\
\enddata
\begin{list}{}{}
\item[Note to the column heads.] $\,$ \\
Name: official meteorite name, \\
Type: classified meteorite type, in which XX-an means anomalous, and metamorphic, \\
Sample ID: from the RELAB database, \\ 
Heated at: experimentally heated temperature (in $^\circ$C), in which the blank indicates unheated, \\
Gr. size: grain size (in $\micron$), \\
$S'$: spectral gradient (in $\%/0.1$ $\micron$). \\
\end{list}
\label{relab}
\end{deluxetable}

\clearpage

\begin{deluxetable}{lcccccc}
 \tabletypesize{\footnotesize}
 \tablewidth{0pt}
 \tablecaption{VIS-NIR spectral features on Phaethon and its meteoritic analog candidates}
 \tablehead{
 \colhead{Sample Name} & \colhead{Heated at} & \colhead{Gr. size} & \colhead{$S'$} & \colhead{UV/$B$ feature} & \colhead{0.7-$\micron$ feature} & \colhead{1.05-$\micron$ feature}\\
 \colhead{} &  \colhead{$^\circ$C} & \colhead{$\micron$} & \colhead{$\%/0.1$ $\micron$} & \colhead{} & \colhead{} & \colhead{}
 }
\startdata
Phaethon            & ($<800$?) &         & $>-11$   & very weak  & no & no  \\
[3pt]
Y-82162 (CI-an)     & ($<750$)  & chip    & $-0.32$  & weak       & no & no  \\
Ivuna (CI1)         &   $700$   & $<125 $ & $-4.24$  & very weak  & no & no  \\
Murchison (CM2)     &    900    & 63--125 & $-0.98$  & weak       & no & no  \\
CK chondrites       &(300--1000)& various & positive & yes        & no & yes \\
\enddata
\begin{list}{}{}
\item[Note.] $\,$ \\
Heated temperature put in parenthesis means estimated temperature. \\
\end{list}
\label{analog}
\end{deluxetable}

\clearpage

\clearpage

\clearpage

\clearpage

\clearpage

\clearpage

\clearpage

\clearpage

\clearpage

\clearpage




\end{document}